\documentclass[]{aa}
\usepackage{graphicx}
\usepackage{natbib}
\usepackage{amssymb}
\usepackage{amsbsy}
\usepackage{txfonts}
\begin{document}
\title{The 10\,$\boldsymbol{\mu}$m amorphous silicate feature of fractal aggregates and compact particles with complex shapes}

\author{M.~Min \and C.~Dominik \and J.~W.~Hovenier \and A.~de~Koter \and L.~B.~F.~M.~Waters}

\institute{Astronomical institute Anton Pannekoek, University of Amsterdam, Kruislaan 403, 1098 SJ  Amsterdam, The Netherlands\\
\email{mmin@science.uva.nl}}
%\date{Received September 15, 1996; accepted March 16, 1997}
\date{Last revision \today}

\abstract{
We model the $10\,\mu$m absorption spectra of nonspherical particles composed of amorphous silicate. We consider two classes of particles, compact ones and fractal aggregates composed of homogeneous spheres. For the compact particles we consider Gaussian random spheres with various degrees of non-sphericity. For the fractal aggregates we compute the absorption spectra for various fractal dimensions. The 10$\,\mu$m spectra are computed for ensembles of these particles in random orientation using the well-known Discrete Dipole Approximation. We compare our results to spectra obtained when using volume equivalent homogeneous spheres and to those computed using a porous sphere approximation. We conclude that, in general, nonspherical particles show a spectral signature that is similar to that of homogeneous spheres with a smaller material volume. This effect is overestimated when approximating the particles by porous spheres with the same volume filling fraction. For aggregates with fractal dimensions typically predicted for cosmic dust, we show that the spectral signature characteristic of very small homogeneous spheres (with a volume equivalent radius $r_V\lesssim 0.5\,\mu$m) can be detected even in very large particles. We conclude that particle sizes are underestimated when using homogeneous spheres to model the emission spectra of astronomical sources. In contrast, the particle sizes are severely overestimated when using equivalent porous spheres to fit observations of 10$\,\mu$m silicate emission.
}

\maketitle

\section{Introduction}
\label{sec:Introduction}

The interpretation of absorption and emission spectra observed from astronomical objects requires knowledge of the absorption cross section as a function of the dust grain characteristics, such as size, shape and composition. Usually the absorption spectra are modeled using homogeneous spherical particles for which calculations can easily be done using Mie theory \citep{Mie}. Although cosmic dust grains are in general not homogeneous spheres, in some cases Mie theory calculations can reproduce the observations quite accurately \citep[see e.g.][]{1974JAtS...31.1137H, 2004ApJ...609..826K}. However, in other cases one has to find a way of modeling the effects of particle shape in order to reproduce observations or laboratory measurements \citep[see e.g.][]{MishHoveTravis}. It is therefore important to know the effects of the adopted particle shape model on the derived dust parameters, such as the particle size and structure \citep[see e.g.][]{2001A&A...378..228F, 2003A&A...404...35M, MinHollow}.

From different formation mechanisms different types of particles may form. For example, when dust grains form from direct gas phase condensation, compact particles may be created. Alternatively, when dust grains stick together to form larger particles, complex aggregated structures may be formed. We study particles in both classes using irregularly shaped compact particles and fractal aggregates. For the compact particle shapes we use so-called Gaussian random spheres \citep{1996JQSRT..55..577M} which allows for a varying degree of irregularity. The fractal aggregates are composed of homogeneous spheres and are constructed using a sequential tunable particle cluster aggregation method, that allows us to construct particles with arbitrary values of the fractal dimension.

We concentrate on the effects of particle non-sphericity on the so-called $10\,\mu$m amorphous silicate feature. Amorphous silicate with an olivine-type stoichiometry is one of the most abundant dust components in various astronomical environments. We will use this type of amorphous silicate and in the rest of this paper refer to it as \emph{amorphous olivine-type silicate}. The absorption spectra of small amorphous silicate grains display a characteristic feature peaking around $10\,\mu$m. The wavelength position of this feature is ideal for ground based observations because of the $10\,\mu$m atmospheric window.

The $10\,\mu$m absorption spectra are computed using the Discrete Dipole Approximation (DDA). 
\citet{1990ApJ...361..251H} proposed to approximate the optical properties of fluffy aggregates with those of so-called equivalent porous spheres. In their method they take a porous sphere with the same volume filling fraction as the fluffy aggregate and compute the optical properties of this porous sphere using effective medium theory. Using this approximation they concluded that the spectral signature of fluffy aggregates is similar to that of compact spheres with a much smaller volume. We test the validity of this approach for calculations of fractal aggregates and complex shaped compact particles by comparing the spectra resulting from this approximate method with those obtained using DDA calculations.
In addition we compare the spectra obtained with DDA to those of volume equivalent homogeneous spheres and analyze the error that is made when employing homogeneous spheres to fit observations of the emission spectra of nonspherical particles.

The method is briefly outlined in Sect.~\ref{sec:method}. The particle shapes are described in Sect.~\ref{sec:shapes}. In Sect.~\ref{sec:results} we present the resulting $10\,\mu$m amorphous silicate spectra. An analysis of our results and a discussion of the implications on the modeling of $10\,\mu$m spectra in astronomical observations is given in Sect.~\ref{sec:discussion}.

\section{Absorption and emission spectra}
\label{sec:method}

The shape of the absorption spectrum of a dust grain, i.e. its absorption cross section as a function of wavelength, contains important information about the dust characteristics. This spectrum can be observed as absorption against a strong infrared background, or as thermal emission.
%In addition, it can be observed from its thermal emission spectrum. The thermal
%emission from an ensemble of isothermal dust grains in random orientation is
%proportional to the orientation averaged absorption cross section times a
%Planck curve with the temperature of the dust grains. Thus we can study the
%shape of the emission spectrum of a dust grain by considering its absorption
%spectrum.

In this paper we consider two measures for the size of a particle, the volume equivalent radius and the circumscribed sphere radius. The volume equivalent radius, $r_V$, is defined as the radius of a sphere with the same material volume as the particle. For very fluffy particles, i.e. particles with a small space filling fraction, the volume equivalent radius does not provide an accurate measure for the actual linear extent of the particle. For this we define the circumscribed sphere radius, $r_c$, to be the radius of the smallest sphere centered on the center of mass of the particle and containing the entire particle.
%maximum distance of all points in the particle to its center of mass. 
As an important parameter we consider the ratio of the circumscribed sphere radius and the volume equivalent radius
\begin{equation}
\label{eq:r_c/r_V}
\gamma=\frac{r_c}{r_V}.
\end{equation}
This parameter determines how dense the material is packed in space. A very fluffy aggregate will have a high value of $\gamma$ while a very dense particle will have a value of $\gamma$ close to unity. For a homogeneous sphere $\gamma=1$. 
%Note that there is a fundamental difference between growing a compact particle
%and growing an aggregate. A compact particle has the same shape, and hence the
%same value of $\gamma$, independent of the size of the particle while an
%aggregate changes shape because extra monomers are added to increase the
%particle size. Thus for an aggregate the value of $\gamma$ can change by
%changing its size.

The mass absorption coefficient, $\kappa$, of a dust grain is defined as its absorption cross section per unit mass
\begin{equation}
\kappa=\frac{C_\mathrm{abs}}{M}\,.
\end{equation}
Here $C_\mathrm{abs}$ is the absorption cross section, and $M$ is the mass of the particle. For grains that are much smaller than the wavelength of radiation both inside and outside the particle, i.e. in the Rayleigh domain, the absorption cross section is proportional to the particle material volume. This means that the mass absorption coefficient of very small particles, which is generally relatively high, is independent of the particle size. For compact particles much larger than the wavelength we are in the geometrical optics domain and the absorption cross section is proportional to the surface area of the particle. For a compact particle this implies that the mass absorption coefficient scales as $r_V^{-1}$ for large values of $r_V$. 
%In general, by increasing the particle size, the mass absorption coefficient
%is decreased.

In general the optical properties of nonspherical particles depend on the orientation of the particle with respect to the incoming electromagnetic field. In the astronomical environments we are interested in, we are dealing with an ensemble of dust grains with random orientations. Therefore, in this paper we will always average the absorption cross section over all particle orientations.

\subsection{The Discrete Dipole Approximation}

The discrete Dipole Approximation (DDA) proposed by \citet{1973ApJ...186..705P} allows for the computation of the optical properties of arbitrarily shaped particles. In the DDA the particle material volume may be divided into small volume elements. Each of these volume elements is assumed to interact with the radiation as a single dipole. The interaction of all dipoles with the incoming field and with the field of all other dipoles is then obtained by solving a $3N\times 3N$ matrix equation, where $N$ is the number of dipoles used to represent the particle volume. For a theoretical foundation of the method in terms of the Maxwell equations see \citet{1992ApJ...394..494L} and \citet{Lakhtakia1993}. For applications of the DDA in the Rayleigh limit, i.e. for sizes of the particles as a whole small compared to the wavelength, see \citet{1993A&A...280..609H, 1995A&A...296..797S}.

In order for the DDA to be valid, the size of the volume elements has to be much smaller than the wavelength of radiation both inside and outside the particle, i.e. \citep[see e.g.][]{1993ApJ...405..685D, 1994OSAJ...11.1491D, Draine2000},
\begin{equation}
\label{eq:DDA}
ka|m|<1,
\end{equation}
where $k=2\pi/\lambda$ is the wavenumber of radiation in vacuum, $\lambda$ the wavelength in vacuum, $a$ the size of the volume element, and $m$ the complex refractive index of the material. 

DDA computations for large particles require a large number of dipoles. Several methods have been suggested to speed up the computations. The most important one makes use of a Fast Fourier Transform (FFT) together with a conjugate gradient solution method \citep[see e.g.][]{Goodman1991, Hoekstra}. A disadvantage of this method is that all dipoles have to be located on a rectangular grid, consequently also the empty, vacuum places on the grid are taken into account in the computation. This slows down the method for fluffy particles with a low volume filling fraction. For example, for particles with a volume filling fraction of 1\%, $N$ is increased by a factor 100.
Therefore, we chose for a direct solution method of the DDA equations. For moderate values of $N$ this allows for relatively fast computations, not only for compact but also for very fluffy grains. An additional advantage is that we only have to do the computation for a particle in a fixed orientation once in order to obtain the optical properties of the particle for any other orientation, whereas the FFT method requires the computation to be repeated for every orientation. Since the computing time required for the direct solution method scales as $N^3$, and the method using FFT scales as $N\log N$, the FFT method becomes faster when large values of $N$ are needed. Typically, both methods are equally fast for $N\sim 10^3$ for a particle with a volume filling fraction of 1\%.

\subsection{Porous spheres}

Another method that is frequently employed to obtain the absorption and emission spectra of aggregated particles is that of approximating the particles by porous spheres \citep{1990ApJ...361..251H, MinHollow, 2005A&A...429..371V}. 
The parameter $\gamma$ can be easily transformed to the so-called \emph{porosity factor}, $P$, which is defined as the vacuum fraction of the volume inside the circumscribing sphere and is given by
\begin{equation}
\label{eq:porosity}
P=1-\frac{r_V^3}{r_c^3}=1-\frac{1}{\gamma^3}\,.
\end{equation}
\citet{1990ApJ...361..251H} use the porosity factor to describe the optical properties of fluffy aggregates. In their approach they take a solid sphere with radius $r_c$ and an effective refractive index depending on the porosity factor. In this way a porous sphere is created with the same volume equivalent radius as well as the same circumscribing sphere radius as the original particle. We will refer to these particles as \emph{equivalent porous spheres}. To compute the effective refractive index \citet{1990ApJ...361..251H} apply the Garnett mixing rule \citep[][usually refered to as the Maxwell-Garnett mixing rule]{1904RSPTA.203..385M}, taking the material as inclusions in a vacuum matrix. Then the effective complex refractive index, $m_\mathrm{eff}$, of a particle with porosity factor $P$ is given by \citep{1990ApJ...361..251H}
\begin{equation}
m_\mathrm{eff}^2=1+\frac{3(1-P)(m^2-1)/(m^2+2)}{1-(1-P)(m^2-1)/(m^2+2)},
\end{equation}
with $m$ the complex refractive index of the material.

The combination of Garnett effective medium approximation with exact Mie calculations for volume equivalent spheres will be referred to as the \emph{porous sphere approximation} throughout this paper.
The validity of this approximation was tested by \citet{1990ApJ...361..251H} for particles where the volume elements are much smaller than the wavelength of incident radiation and in addition are randomly distributed over the circumscribing sphere volume. They found that in this case, the porous sphere approximation yields rather accurate results. However, in realistic aggregates the constituents are not randomly distributed in space. In contrast, the constituents have to be touching to form, for example, fractal like structures. We will test the validity of the porous sphere approximation for computations of emission spectra of fractal aggregates and complex shaped compact particles by comparing the spectra resulting from this approximation with those obtained from DDA computations.

\section{Particle shapes}
\label{sec:shapes}

The shape of a dust grain is an important parameter determining its spectroscopic characteristics. In this paper we distinguish between two classes of shapes, aggregates and compact particles. For the compact particles we use Gaussian random spheres, while for the aggregates we use fractal aggregates with varying fractal dimensions.

\subsection{Gaussian Random Spheres}

\begin{figure*}[!t]
\begin{center}
\def\setw{3.0cm}
\def\sett{1.2cm}
\resizebox{15cm}{!}{
\begin{tabular}{cccc}
$\sigma=0.1$ & $\sigma=0.3$ & $\sigma=0.5$ & $\sigma=0.7$	\\
\parbox{\setw}{\includegraphics[width=\setw]{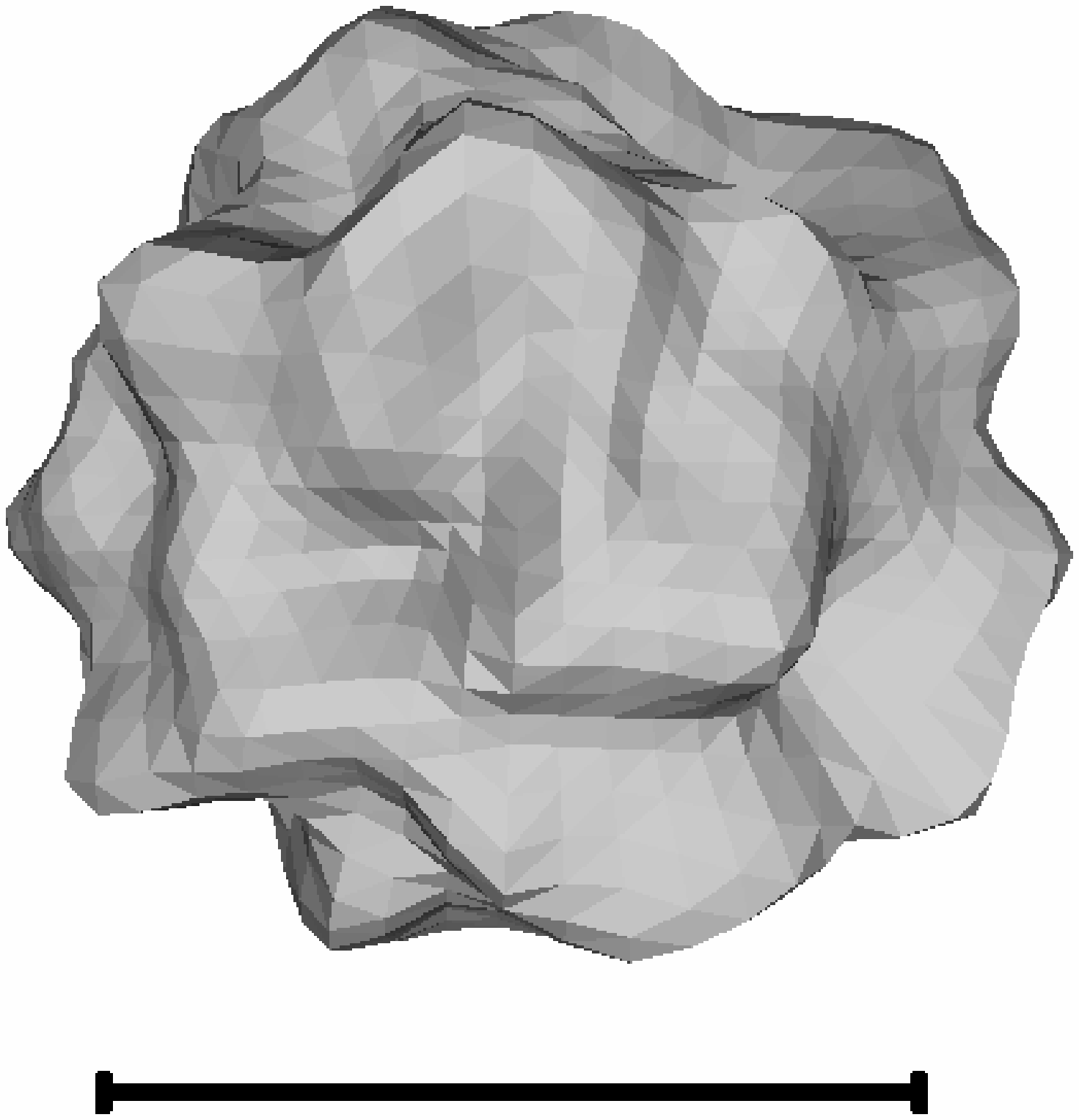}\\
\centerline{$\gamma=1.32$}} &
\parbox{\setw}{\includegraphics[width=\setw]{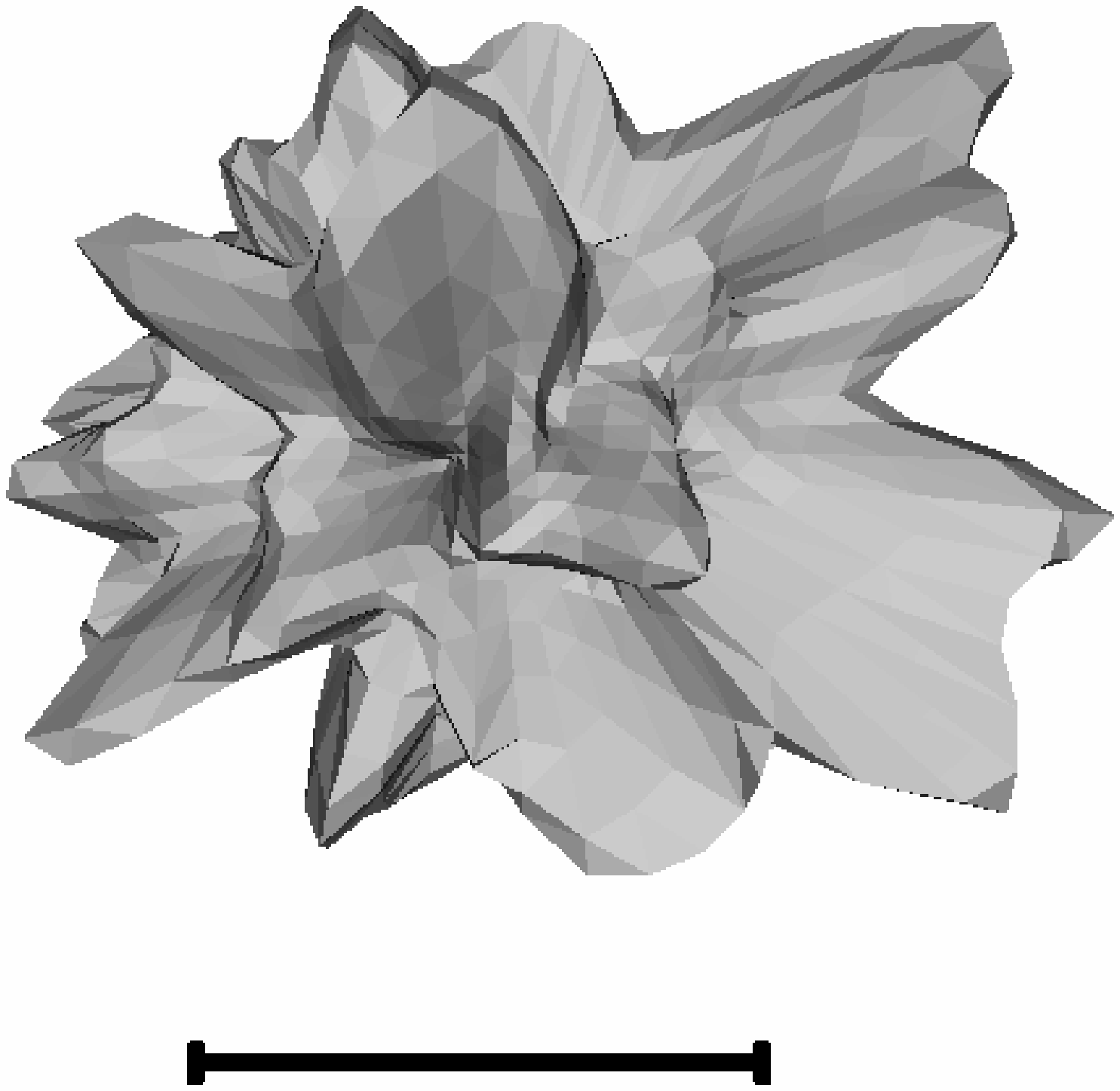}\\
\centerline{$\gamma=2.00$}} &
\parbox{\setw}{\includegraphics[width=\setw]{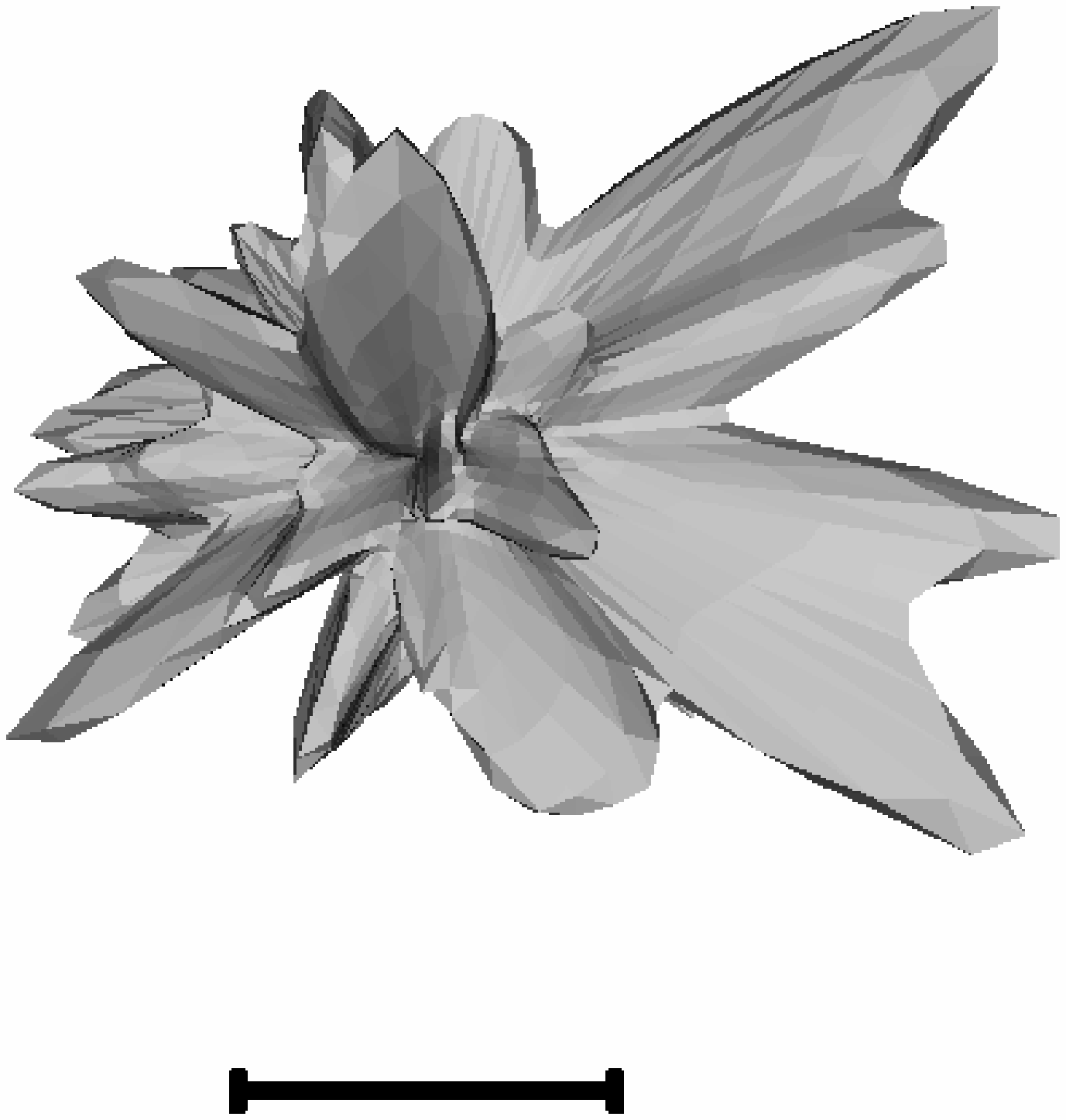}\\
\centerline{$\gamma=2.64$}} &
\parbox{\setw}{\includegraphics[width=\setw]{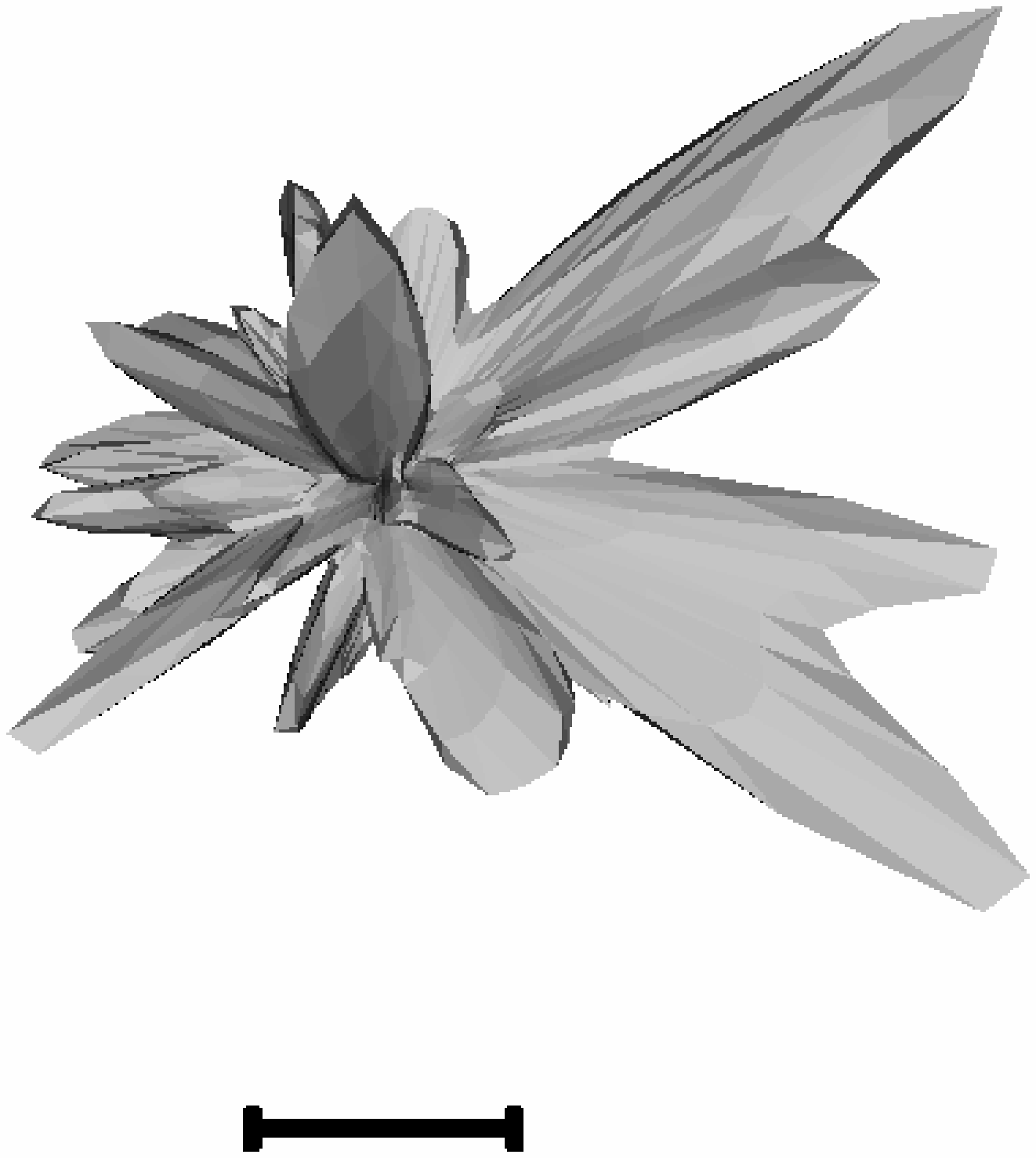}\\
\centerline{$\gamma=3.06$}} \\
\end{tabular}
}
\end{center}
\caption{Pictures of the Gaussian random spheres employed in our calculations. The diameters of the corresponding volume equivalent spheres are indicated by bars below the particles.}
\label{fig:Gaussian}
\end{figure*}

%In order to model the properties of complex shaped compact particles we have to
%employ a particle shape model. Here we use Gaussian random spheres. 
A Gaussian random sphere \citep{1996JQSRT..55..577M} is a homogeneous sphere the surface of which is distorted according to a Gaussian random distribution. The distortion of the surface is parameterized by two shape variables, the standard deviation of the distance to the center, $\sigma$, and the average correlation angle, $\Gamma$. Using various values for these two parameters allows us to construct particles with varying degrees of irregularity. The number of `hills and valleys' on the surface in a solid angle is determined by the value of $\Gamma$, while $\sigma$ determines the height of these hills and valleys. For details see \citet{1996JQSRT..55..577M}. 

In Fig.~\ref{fig:Gaussian} we show pictures of the Gaussian random spheres used in our computations. We vary the value of $\sigma$ from $0.1$ to $0.7$ and fix the value of $\Gamma$ to $10^\circ$. 
%Since the Gaussian random spheres are compact particles, changing the size 
%only scales the entire particle. Therefore, $\gamma$ is the same for all 
%particle sizes.

\subsection{Fractal aggregates}

In astronomical environments large grains form by aggregation of small particles \citep[see e.g.][]{kempf99}. Studies on interplanetary dust particles from likely cometary origin show that cometary dust grains are likely to be aggregates. A special class of aggregates are the so-called fractal aggregates. 
%In nature many coagulation mechanisms can produce fractal aggregates. The 
%fractal dimension of these aggregates depends on the conditions in the 
%environment in which they form.

A fractal aggregate composed of homogeneous spheres obeys the scaling law \citep{Filippov}
\begin{equation}
\label{eq:scaling law}
N=k_f\left(\frac{R_g}{a}\right)^{D_f}.
\end{equation}
Here $N$ is the number of constituents each with radius $a$; $k_f$ is the fractal prefactor; $D_f$ is the fractal dimension, and $R_g$ is the radius of gyration defined by
\begin{eqnarray}
R_g^2&=&\frac{1}{N}\sum_{i=1}^N \left|\vec{r}_i-\vec{r}_0\right|^2,\\
\vec{r}_0&=&\frac{1}{N}\sum_{i=1}^N \vec{r}_i\,,
\end{eqnarray}
where $\vec{r}_i$ is the position of the $i$th constituent, and $\vec{r}_0$ is the center of mass.
The value of the fractal dimension can in theory vary between the two extremes $D_f=1$ (a thin, straight chain of particles) and $D_f=3$ (a homogeneous sphere).

We use a sequential tunable particle-cluster aggregation method developed by \citet{Filippov} to construct the fractal aggregates. This method allows us to construct aggregates with arbitrary fractal dimension. 

\begin{figure*}[!t]
\begin{center}
\def\setw{3.0cm}
\def\sett{1.2cm}
\resizebox{15cm}{!}{
\begin{tabular}{c|cccc}
	&	$D_f=2.8$	&	$D_f=2.4$	&	$D_f=1.8$	&	$D_f=1.2$	\\
\hline\vspace{-0.3cm}\\
\parbox{\sett}{\rotatebox{90}{\parbox{\setw}{\begin{center}$r_V=0.4\,\mu$m\\N = 1\end{center}}}}&
\parbox{\setw}{\includegraphics[width=\setw]{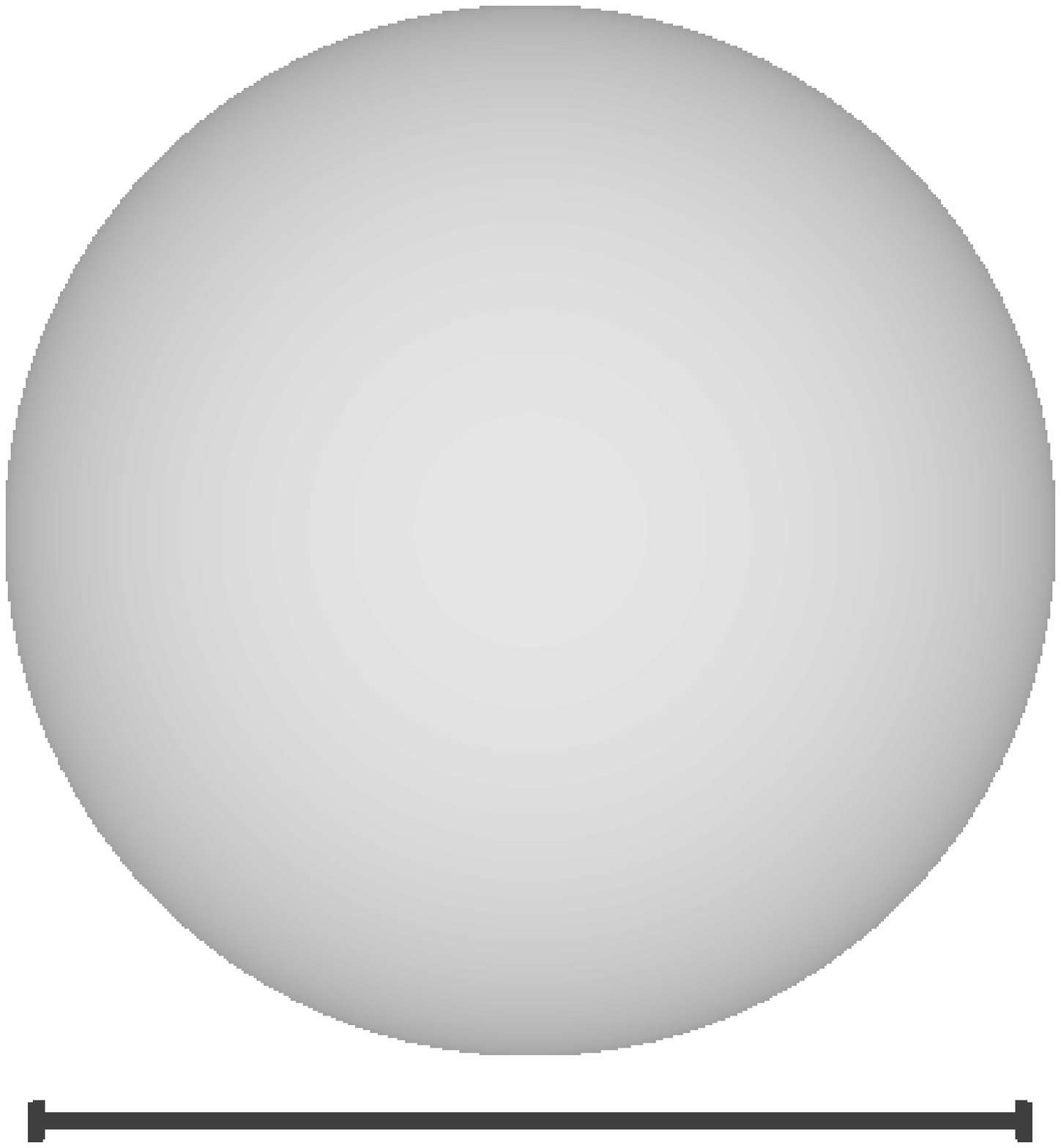}\\
\centerline{$\gamma=1.0$}} &
\parbox{\setw}{\includegraphics[width=\setw]{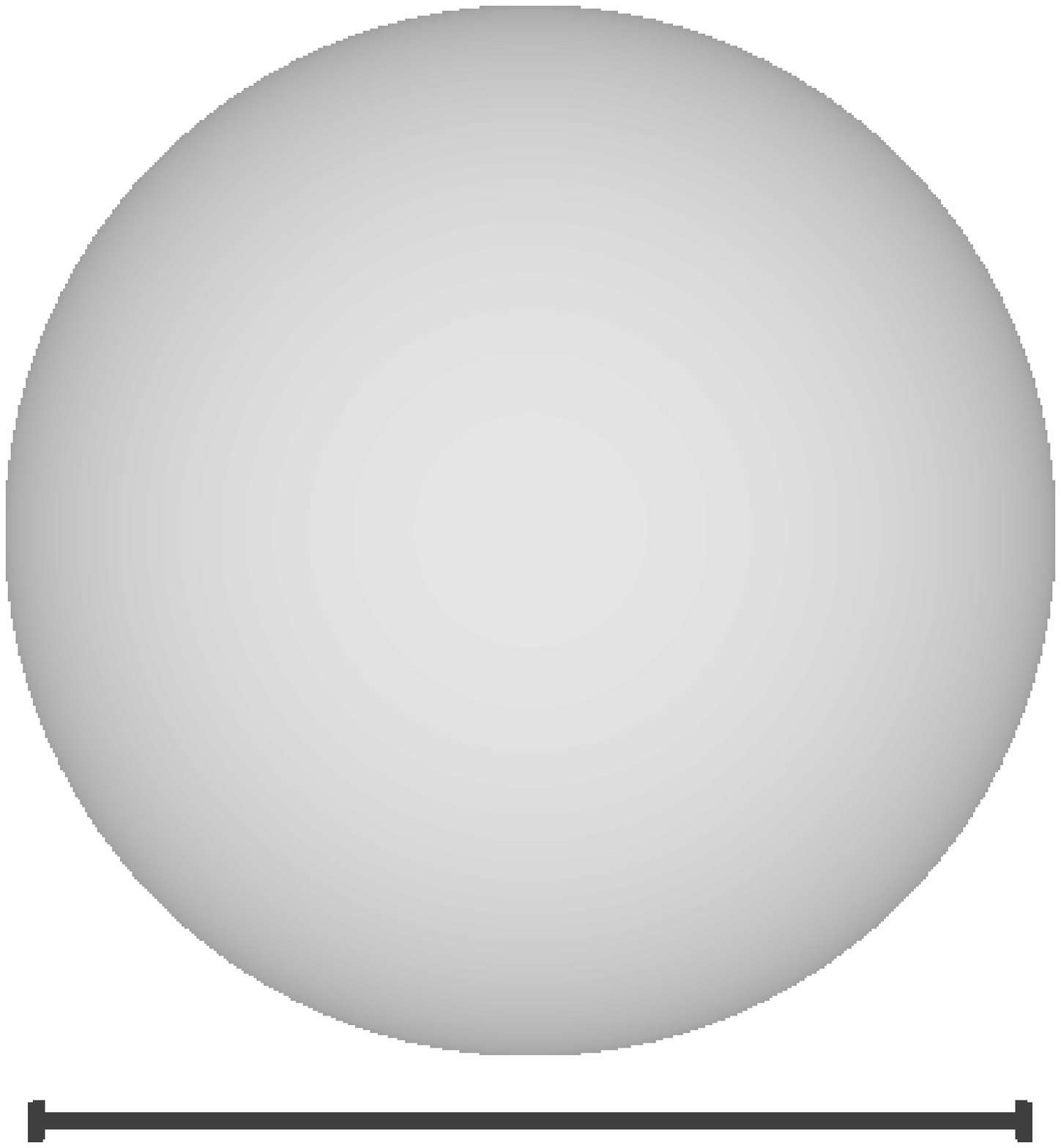}\\
\centerline{$\gamma=1.0$}} &
\parbox{\setw}{\includegraphics[width=\setw]{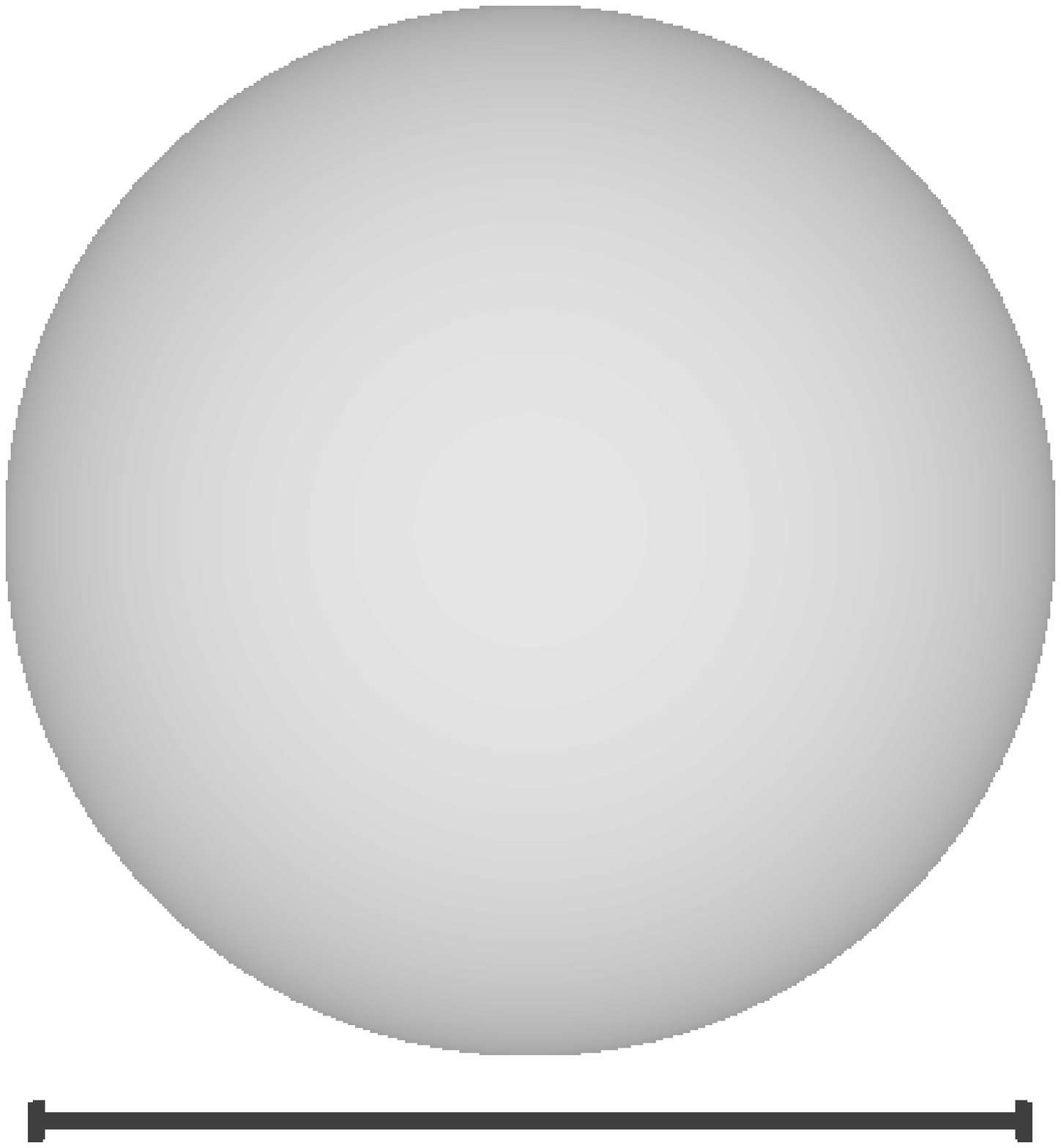}\\
\centerline{$\gamma=1.0$}} &
\parbox{\setw}{\includegraphics[width=\setw]{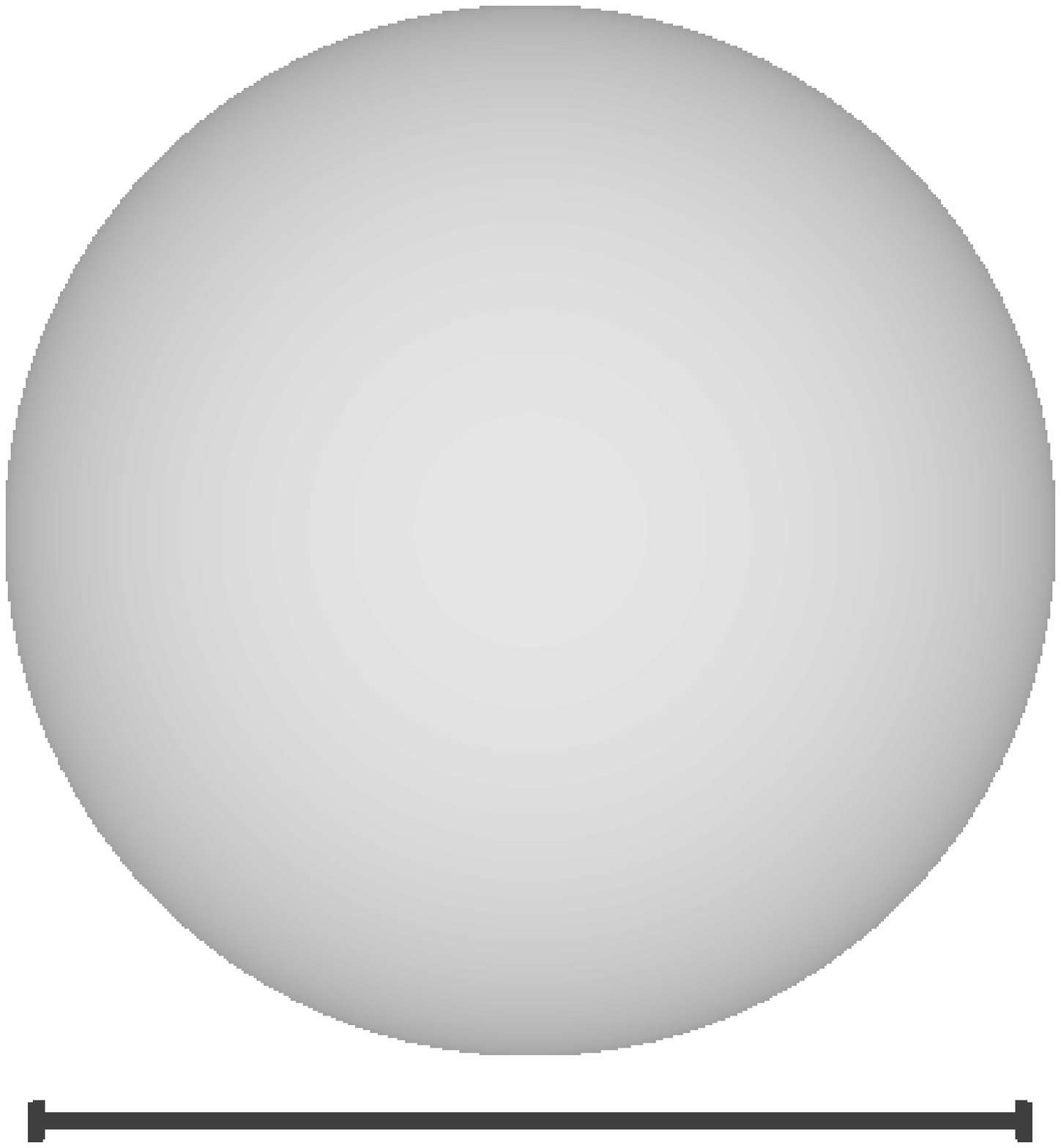}\\
\centerline{$\gamma=1.0$}} \\
\vspace{-0.3cm}\\
\hline\vspace{-0.3cm}\\
\parbox{\sett}{\rotatebox{90}{\parbox{\setw}{\begin{center}$r_V=1.0\,\mu$m\\N = 16\end{center}}}}&
\parbox{\setw}{\includegraphics[width=\setw]{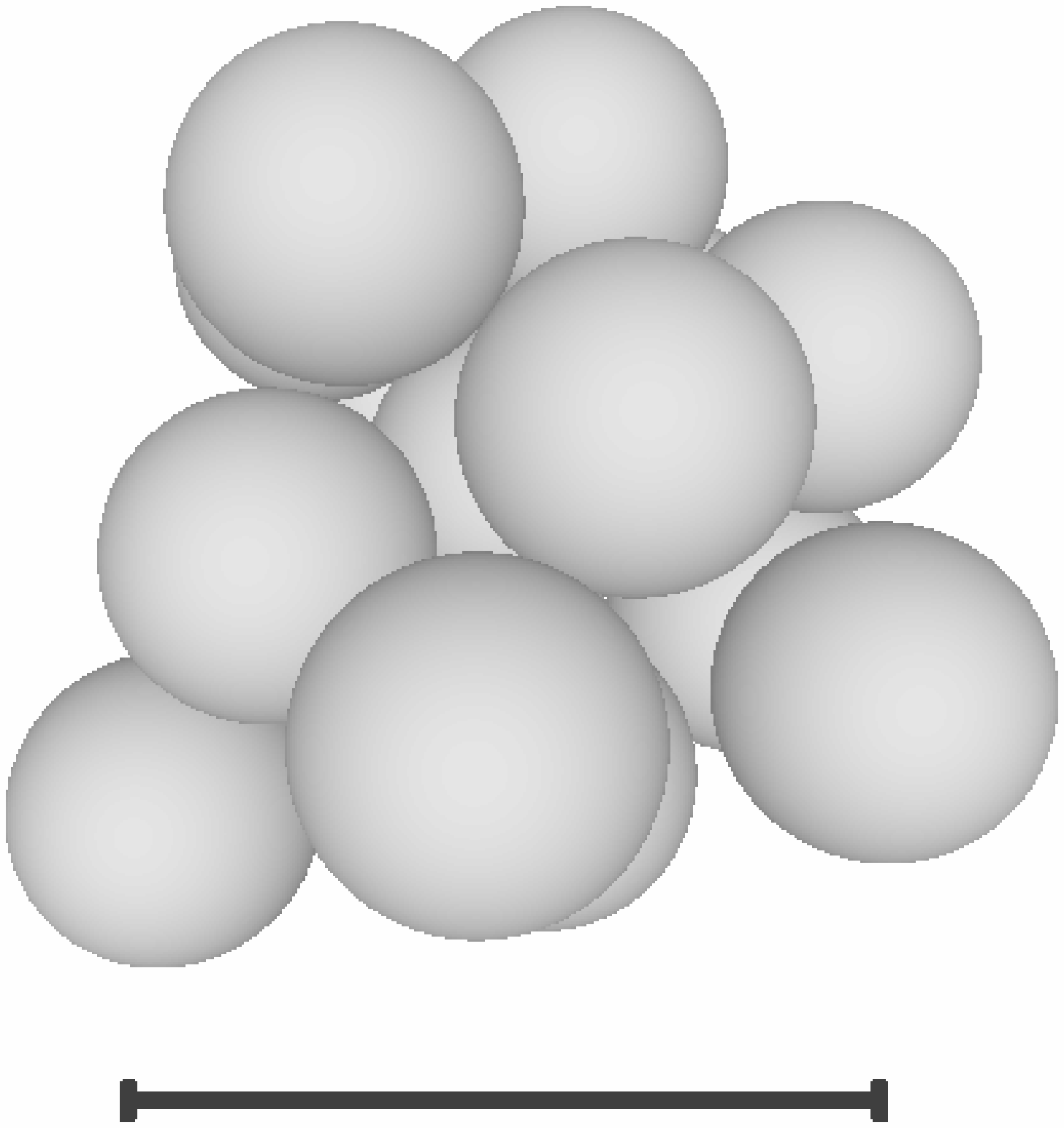}\\
\centerline{$\gamma=1.57$}} &
\parbox{\setw}{\includegraphics[width=\setw]{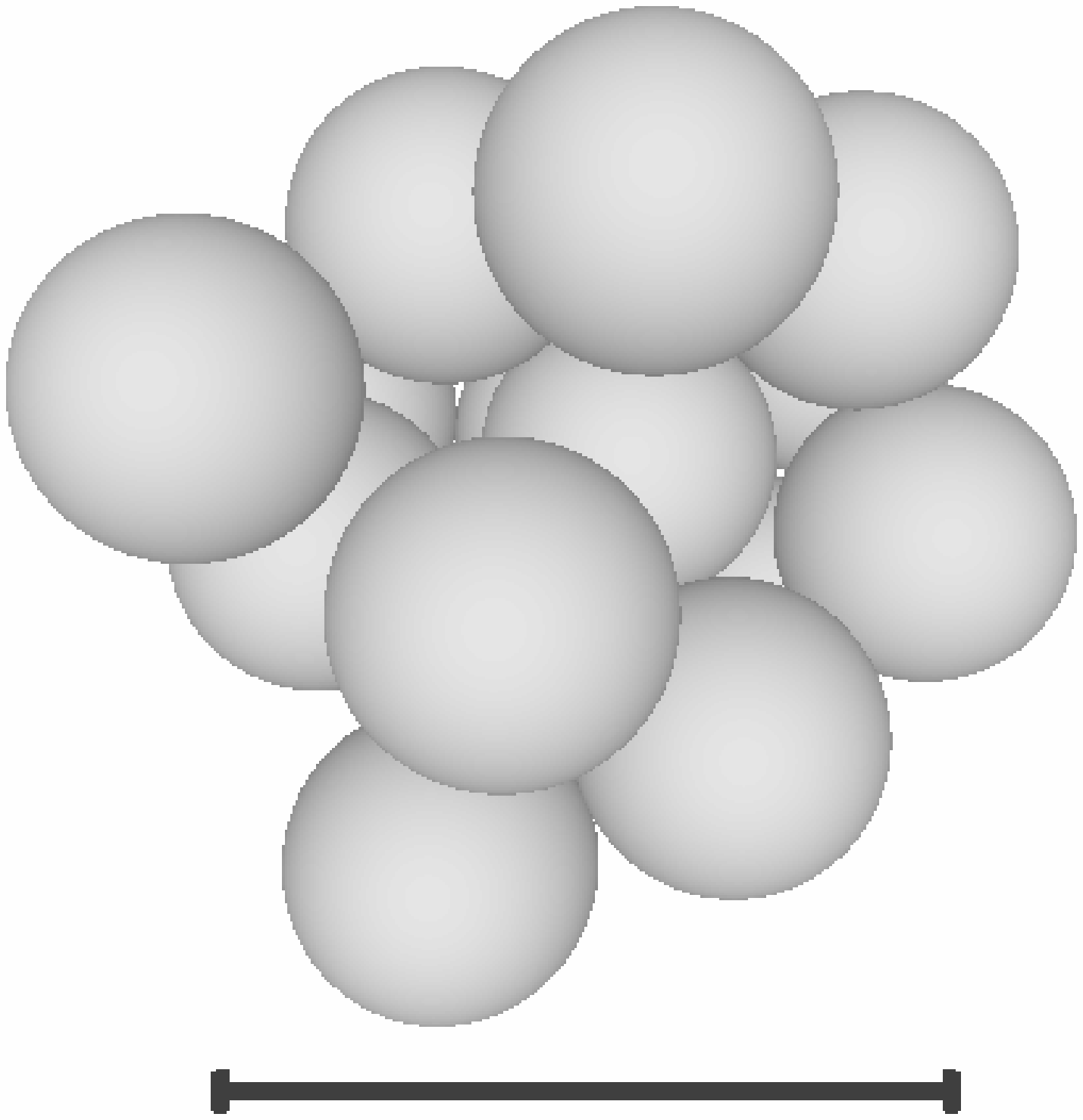}\\
\centerline{$\gamma=1.56$}} &
\parbox{\setw}{\includegraphics[width=\setw]{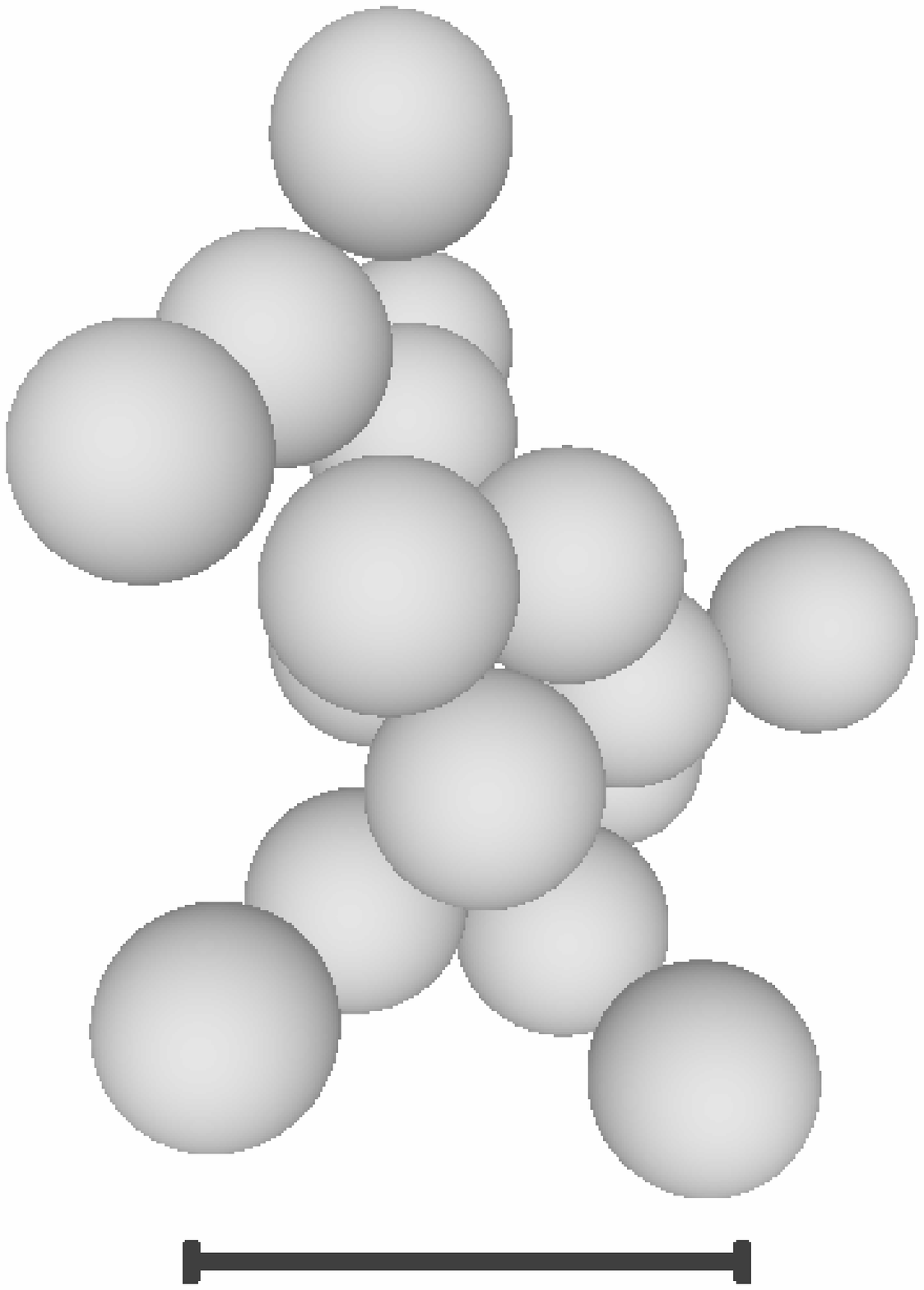}\\
\centerline{$\gamma=2.13$}} &
\parbox{\setw}{\includegraphics[width=\setw]{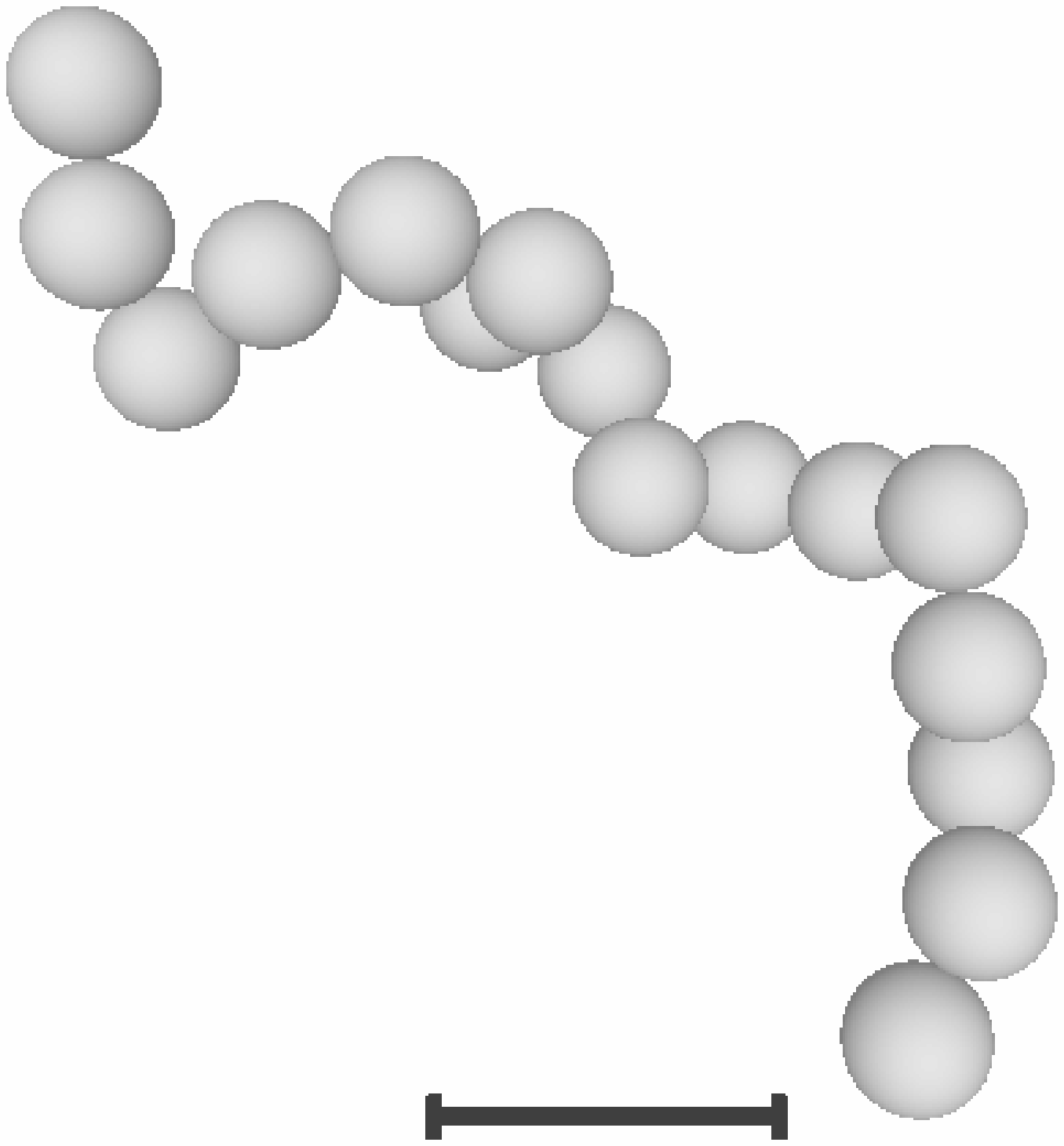}\\
\centerline{$\gamma=3.84$}} \\
\vspace{-0.3cm}\\
\hline\vspace{-0.3cm}\\
\parbox{\sett}{\rotatebox{90}{\parbox{\setw}{\begin{center}$r_V=2.0\,\mu$m\\N = 125\end{center}}}}&
\parbox{\setw}{\includegraphics[width=\setw]{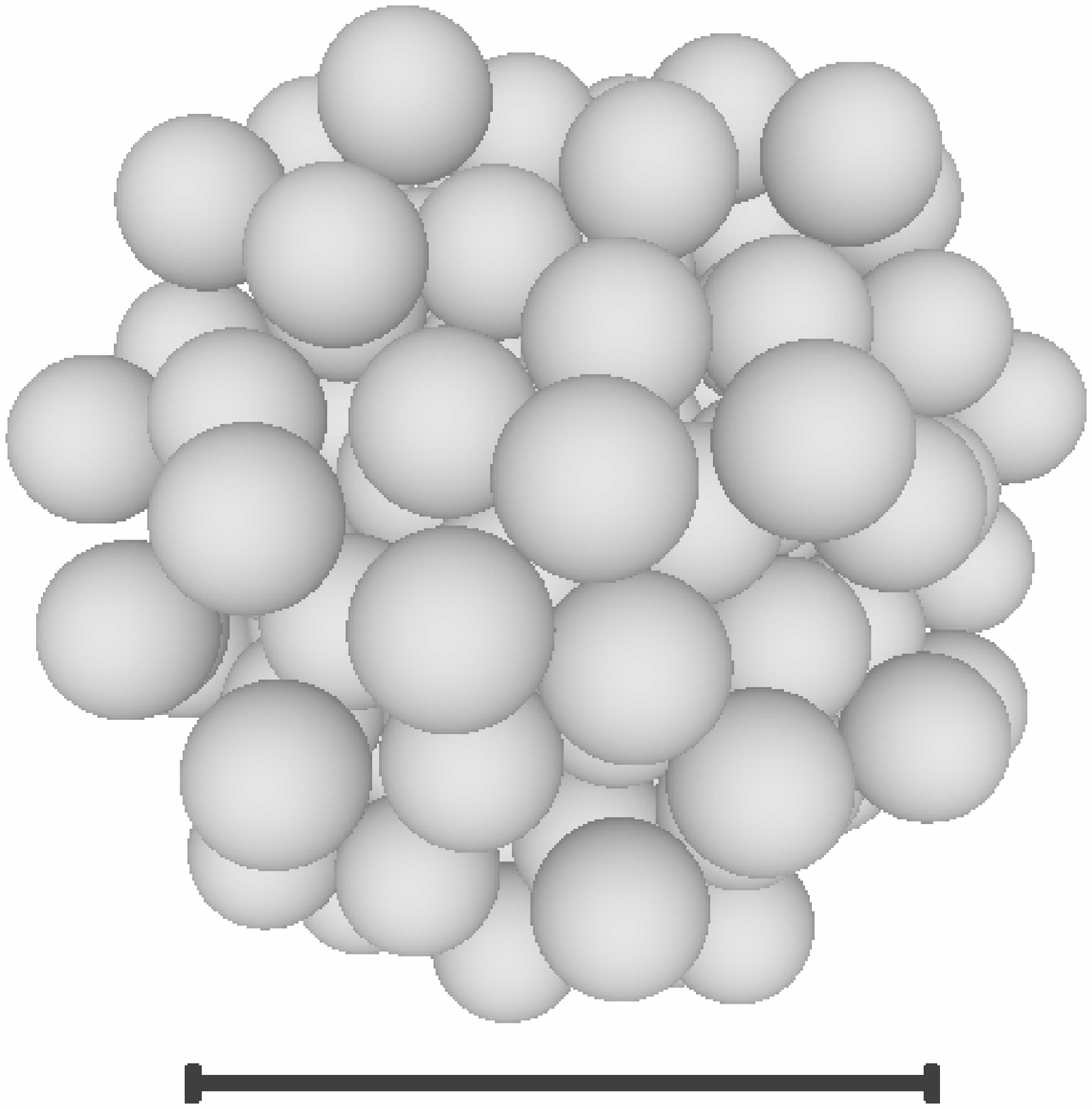}\\
\centerline{$\gamma=1.44$}} &
\parbox{\setw}{\includegraphics[width=\setw]{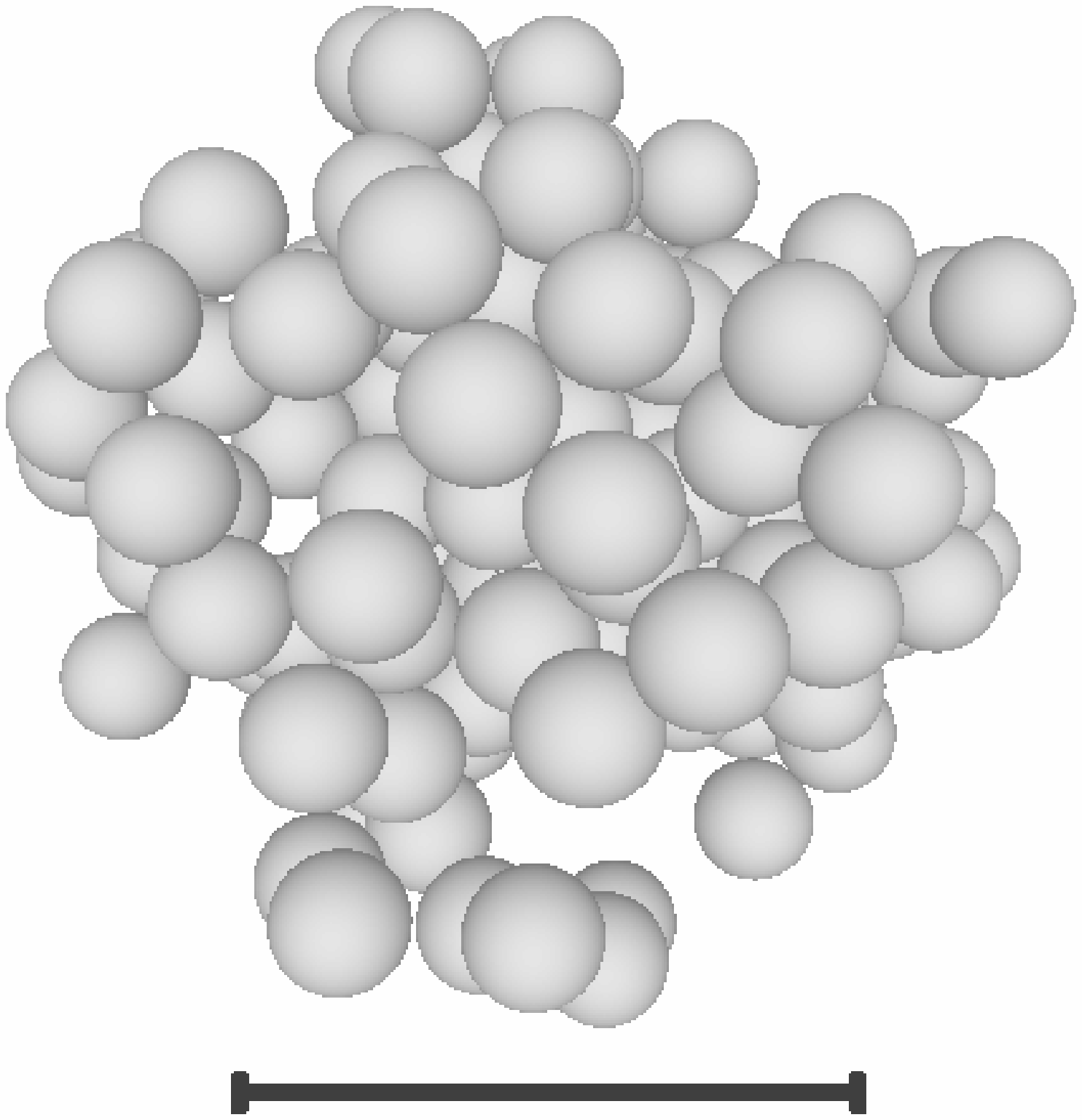}\\
\centerline{$\gamma=1.71$}} &
\parbox{\setw}{\includegraphics[width=\setw]{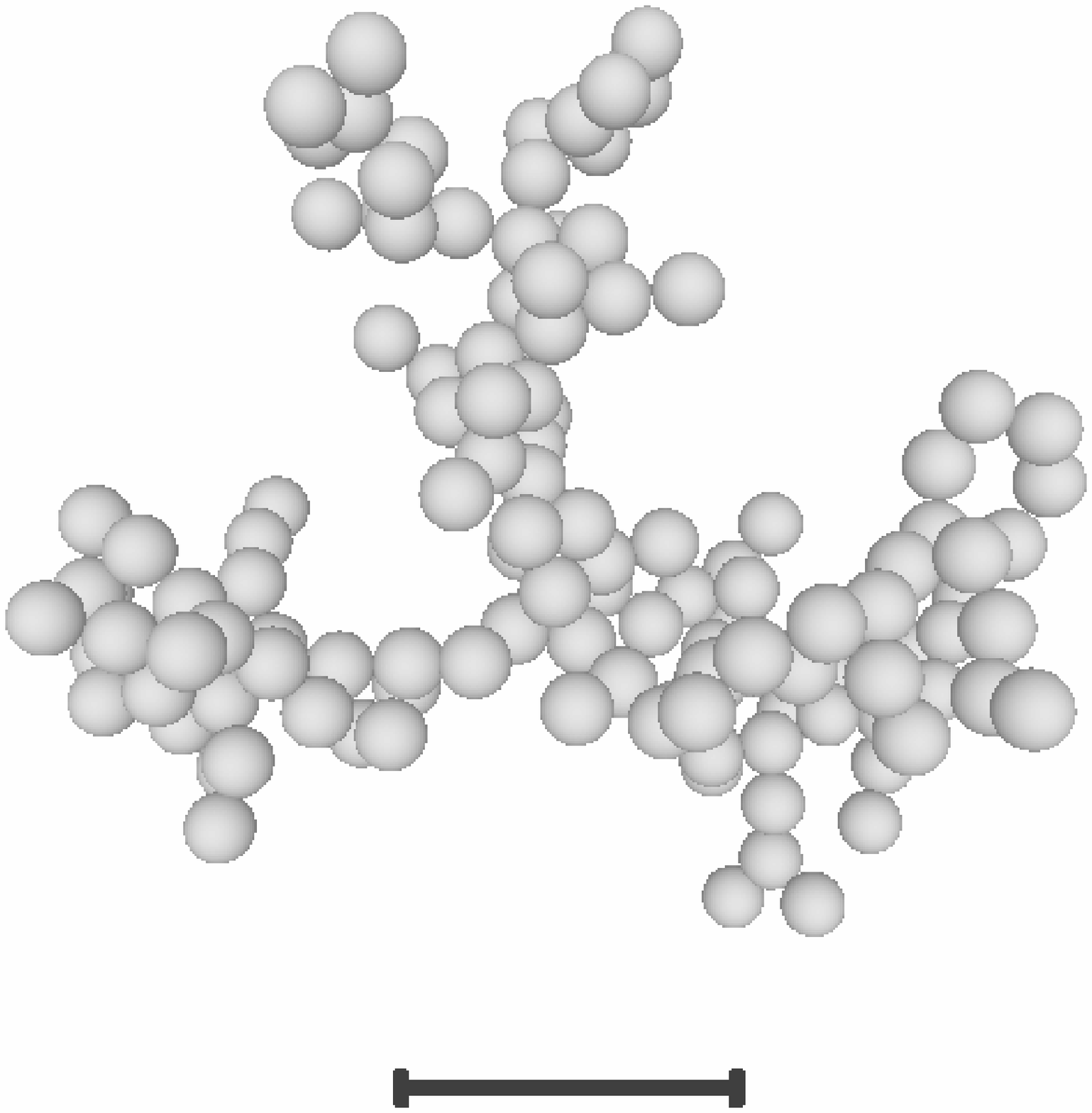}\\
\centerline{$\gamma=3.07$}} &
\parbox{\setw}{\includegraphics[width=\setw]{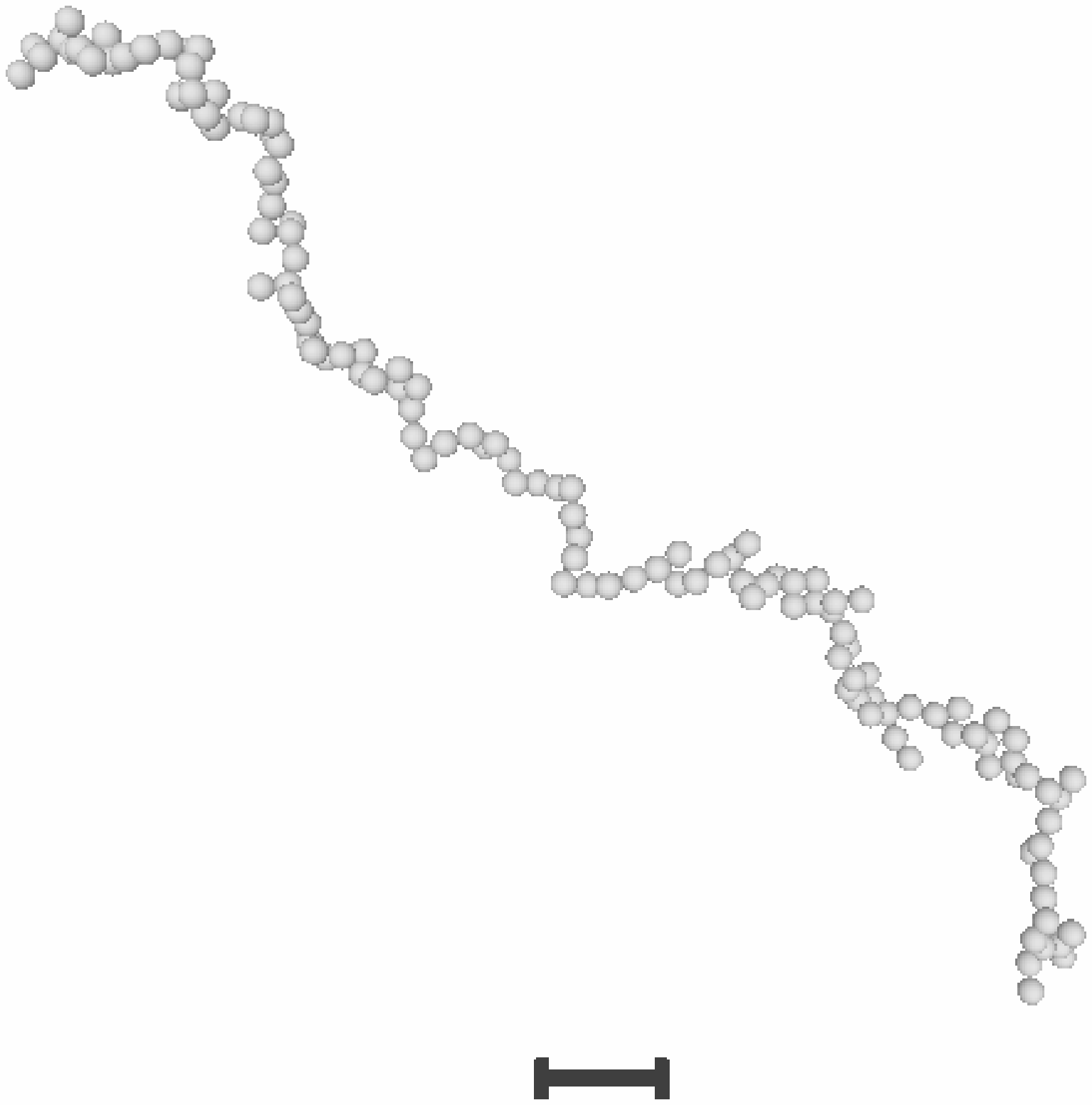}\\
\centerline{$\gamma=10.4$}} \\
\vspace{-0.3cm}\\
\hline\vspace{-0.3cm}\\
\parbox{\sett}{\rotatebox{90}{\parbox{\setw}{\begin{center}$r_V=4.0\,\mu$m\\N = 1000\end{center}}}}&
\parbox{\setw}{\includegraphics[width=\setw]{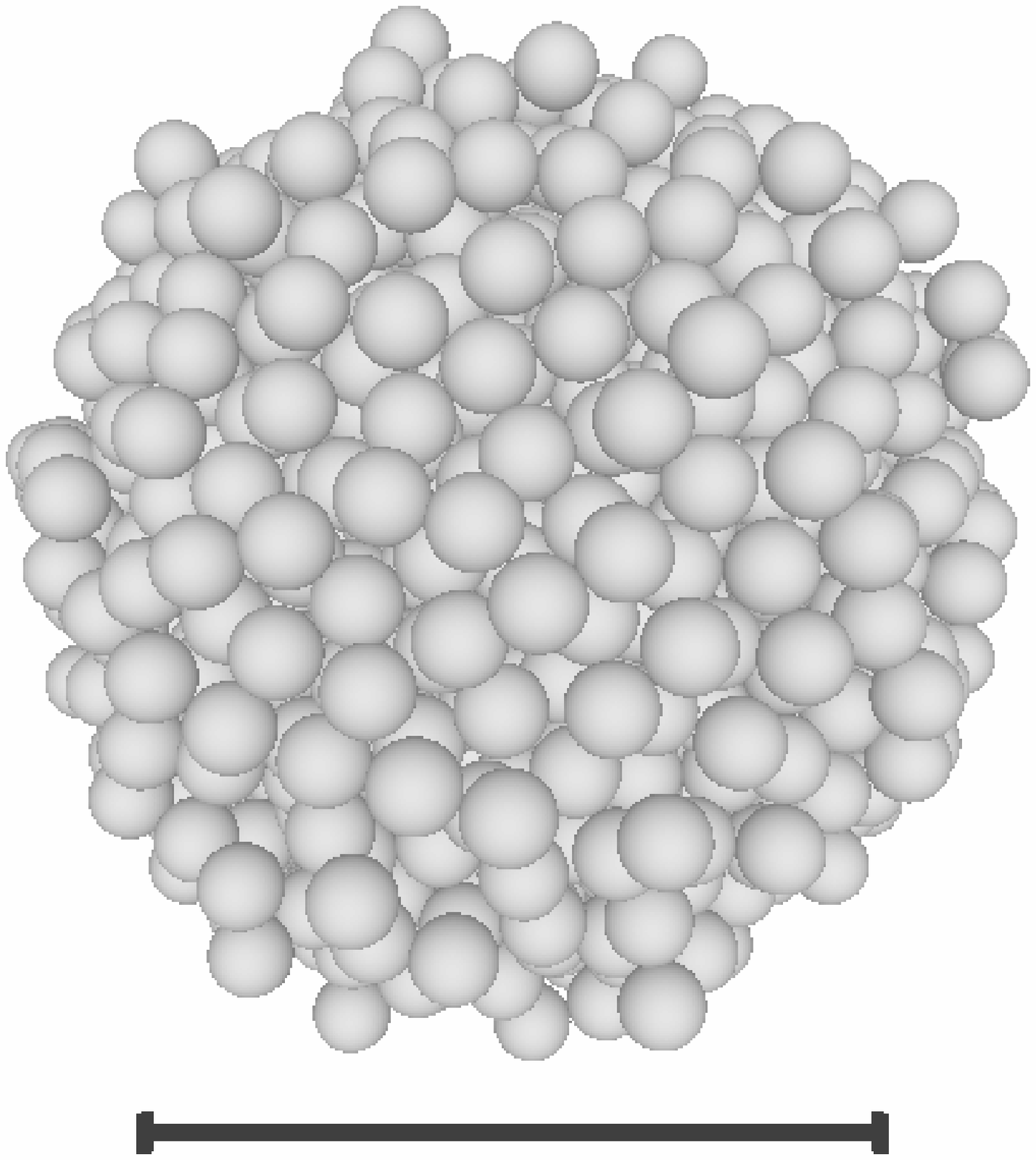}\\
\centerline{$\gamma=1.42$}} &
\parbox{\setw}{\includegraphics[width=\setw]{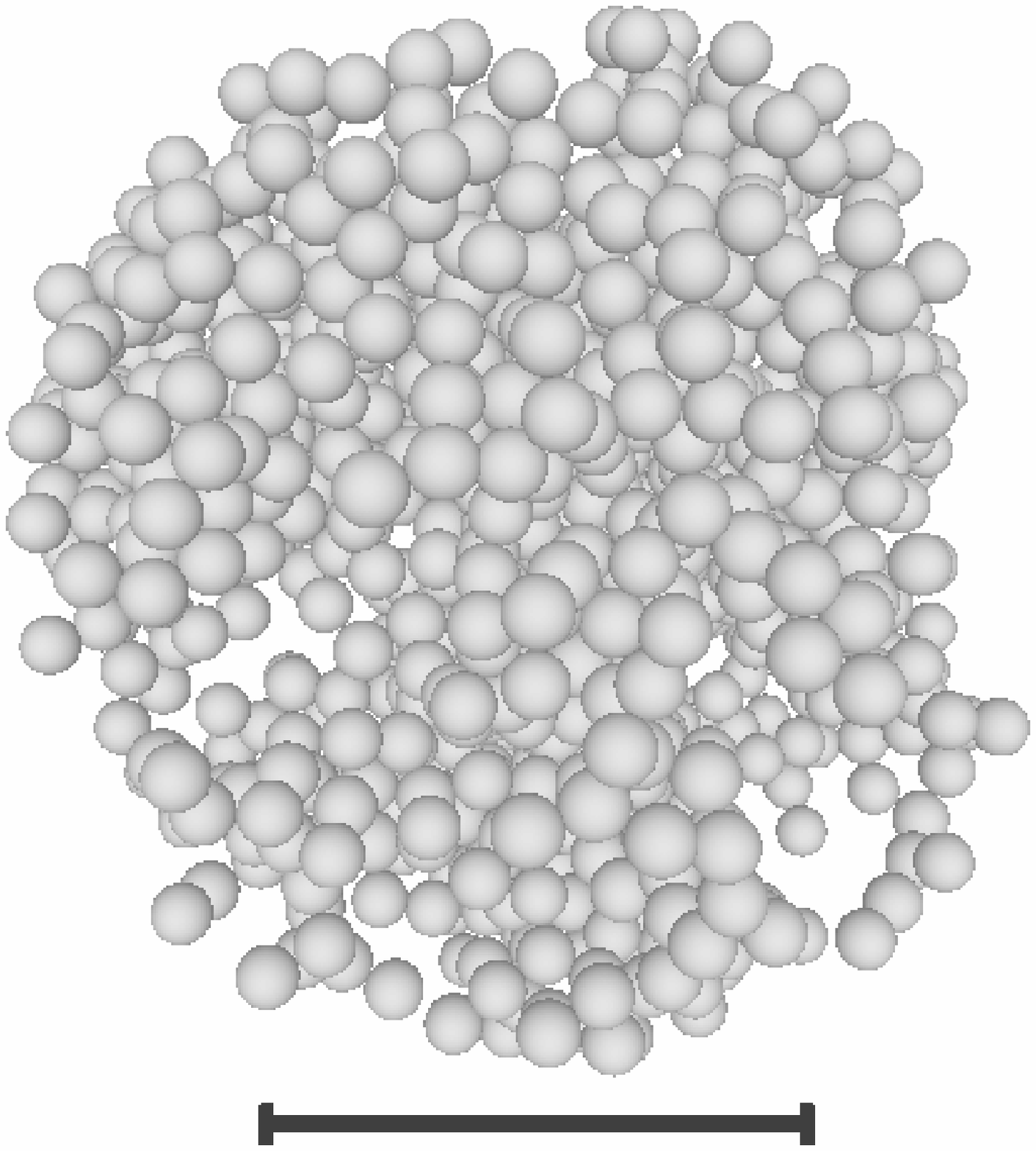}\\
\centerline{$\gamma=1.90$}} &
\parbox{\setw}{\includegraphics[width=\setw]{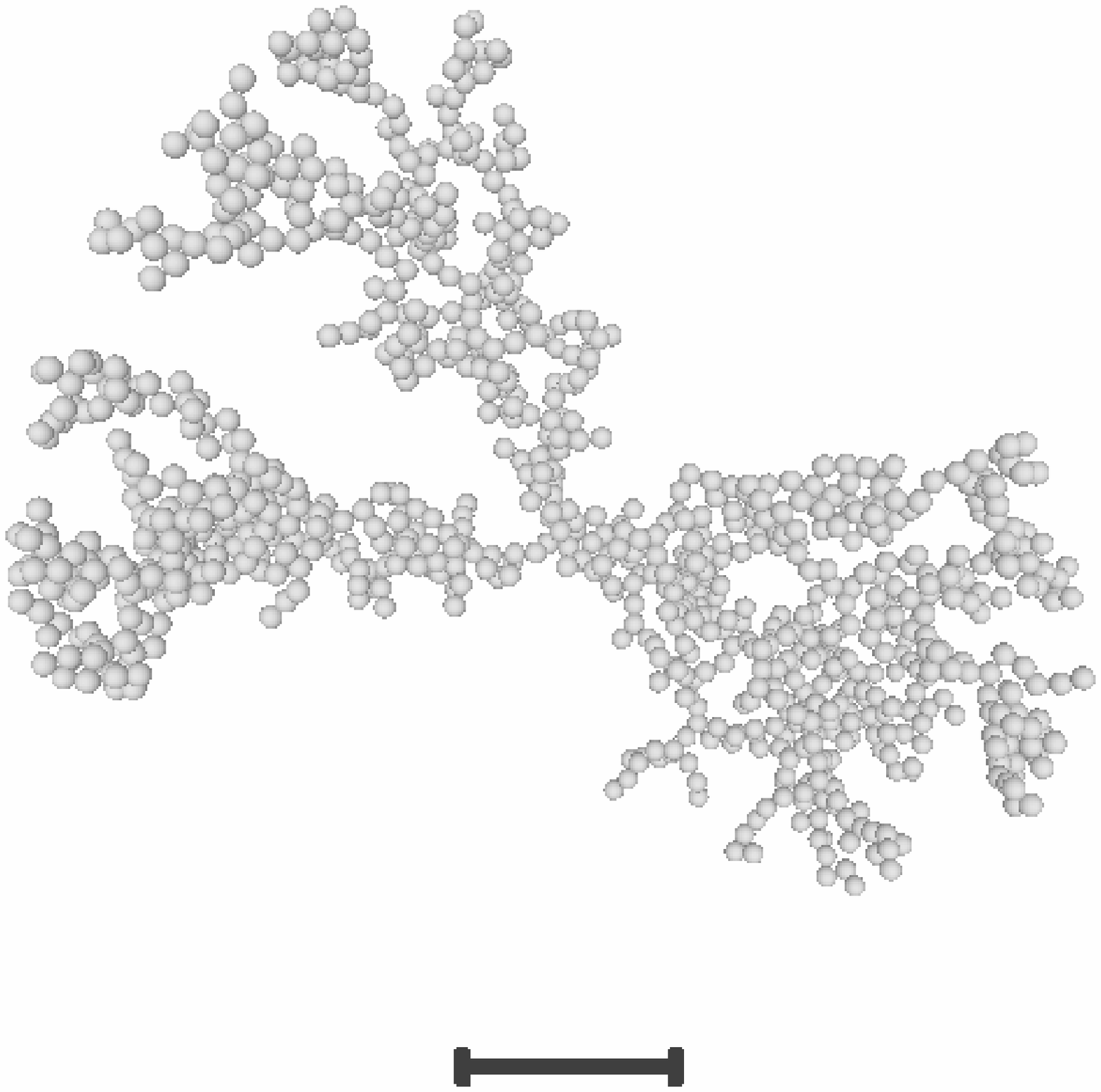}\\
\centerline{$\gamma=4.69$}} &
\parbox{\setw}{\includegraphics[width=\setw]{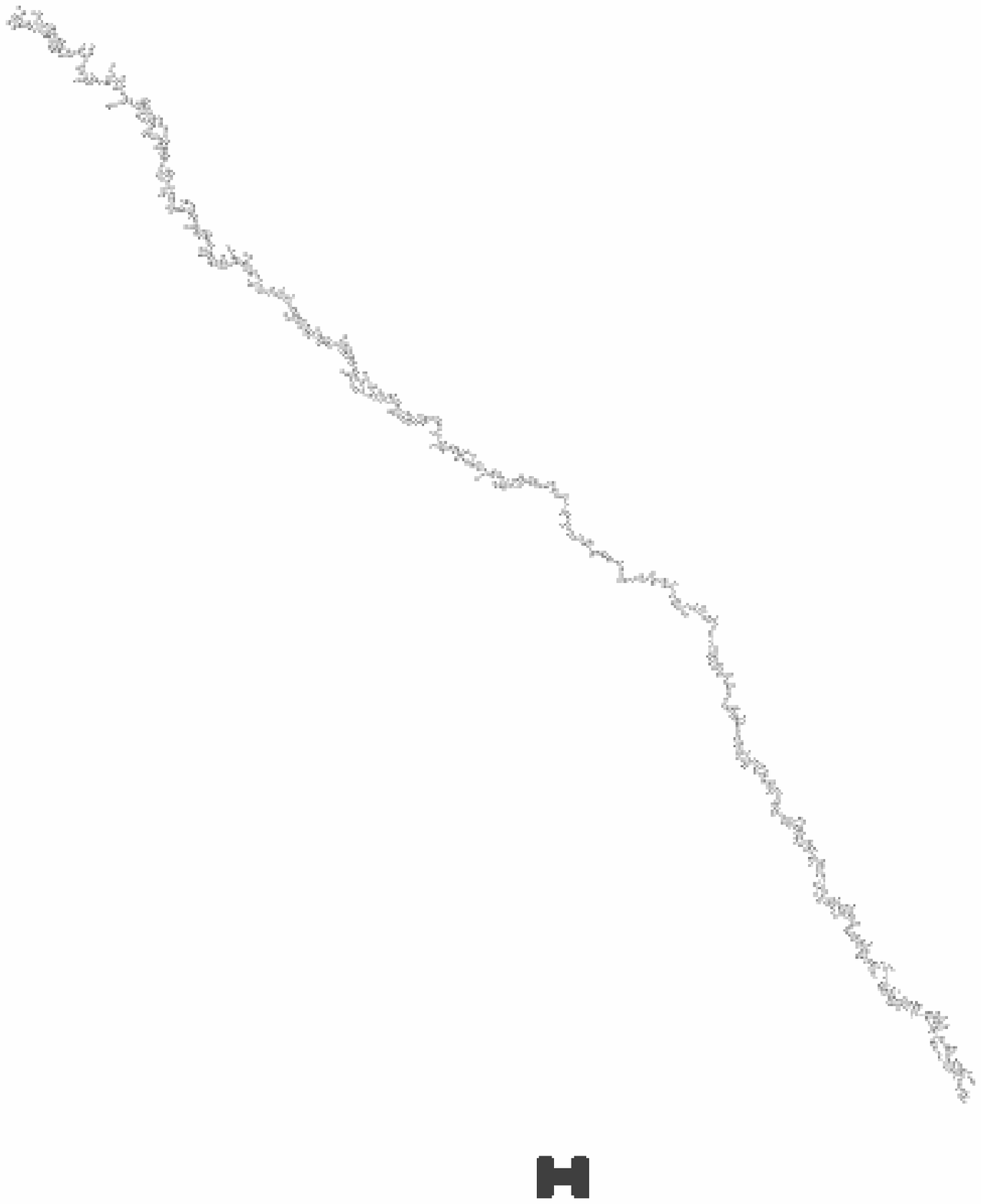}\\
\centerline{$\gamma=29.1$}} \\
\vspace{-0.3cm}\\
\hline\vspace{-0.3cm}\\
\parbox{\sett}{\rotatebox{90}{\parbox{\setw}{\begin{center}$r_V=6.0\,\mu$m\\N = 3375\end{center}}}}&
\parbox{\setw}{\includegraphics[width=\setw]{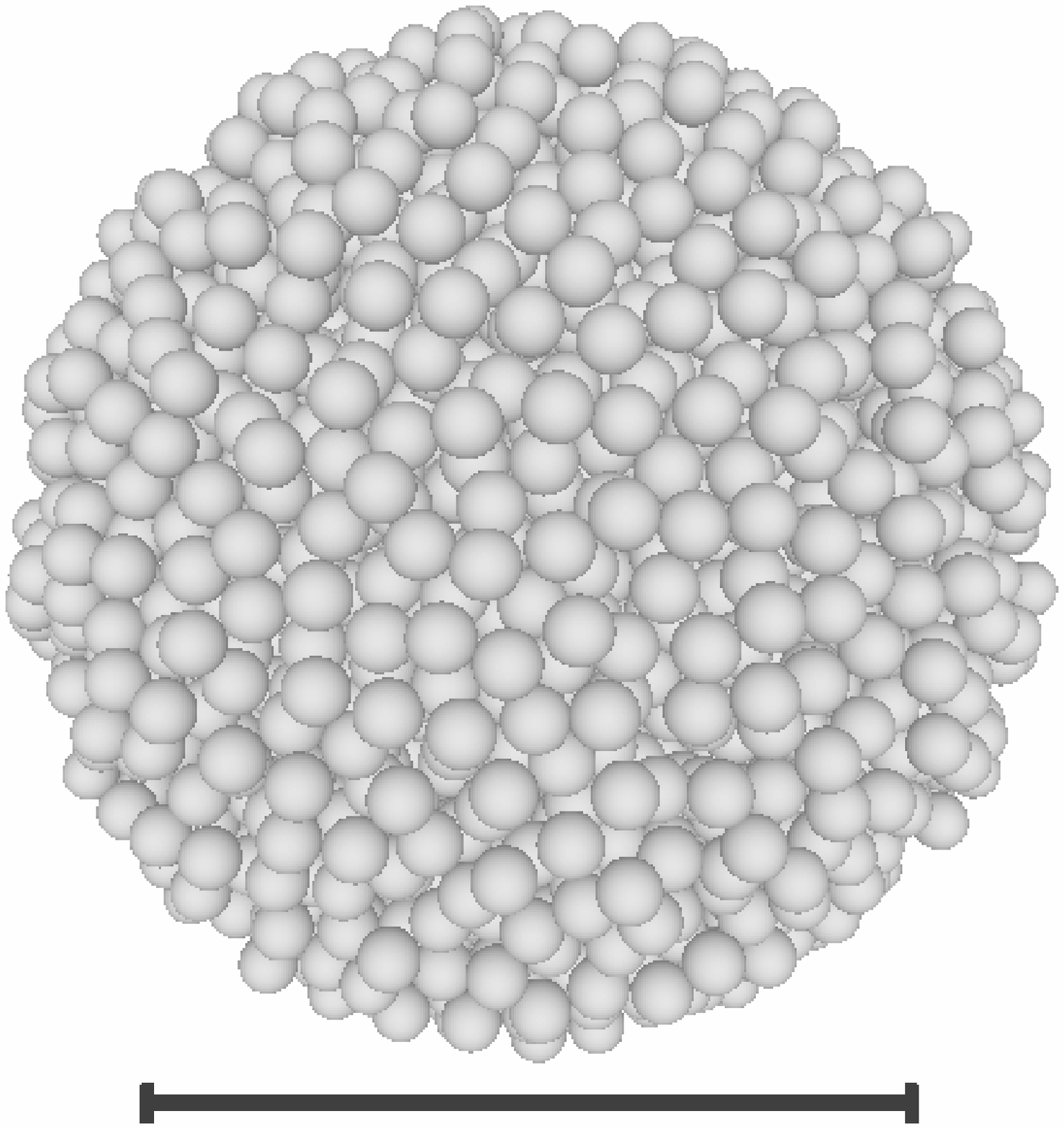}\\
\centerline{$\gamma=1.31$}} &
\parbox{\setw}{\includegraphics[width=\setw]{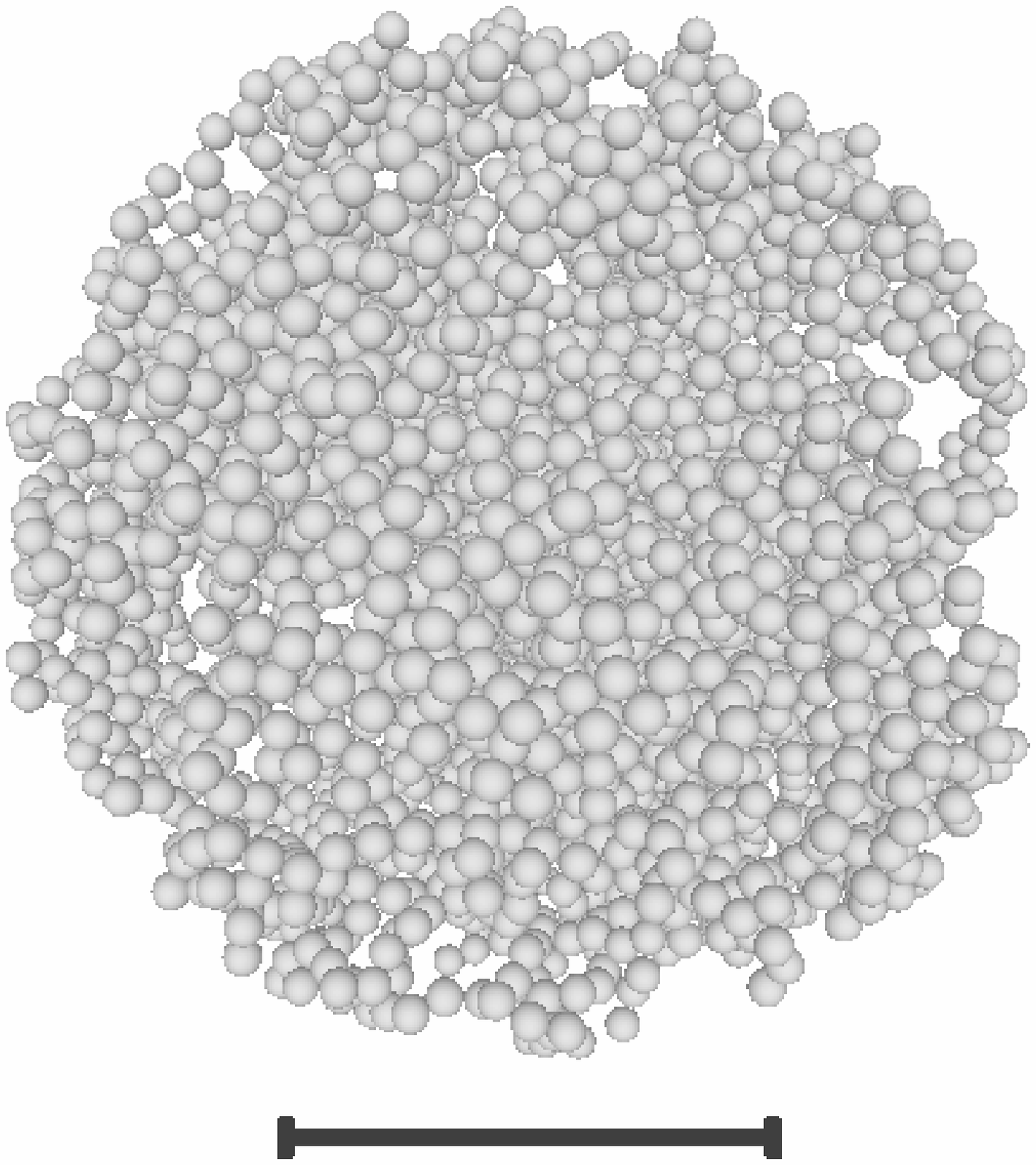}\\
\centerline{$\gamma=2.06$}} &
\parbox{\setw}{\includegraphics[width=\setw]{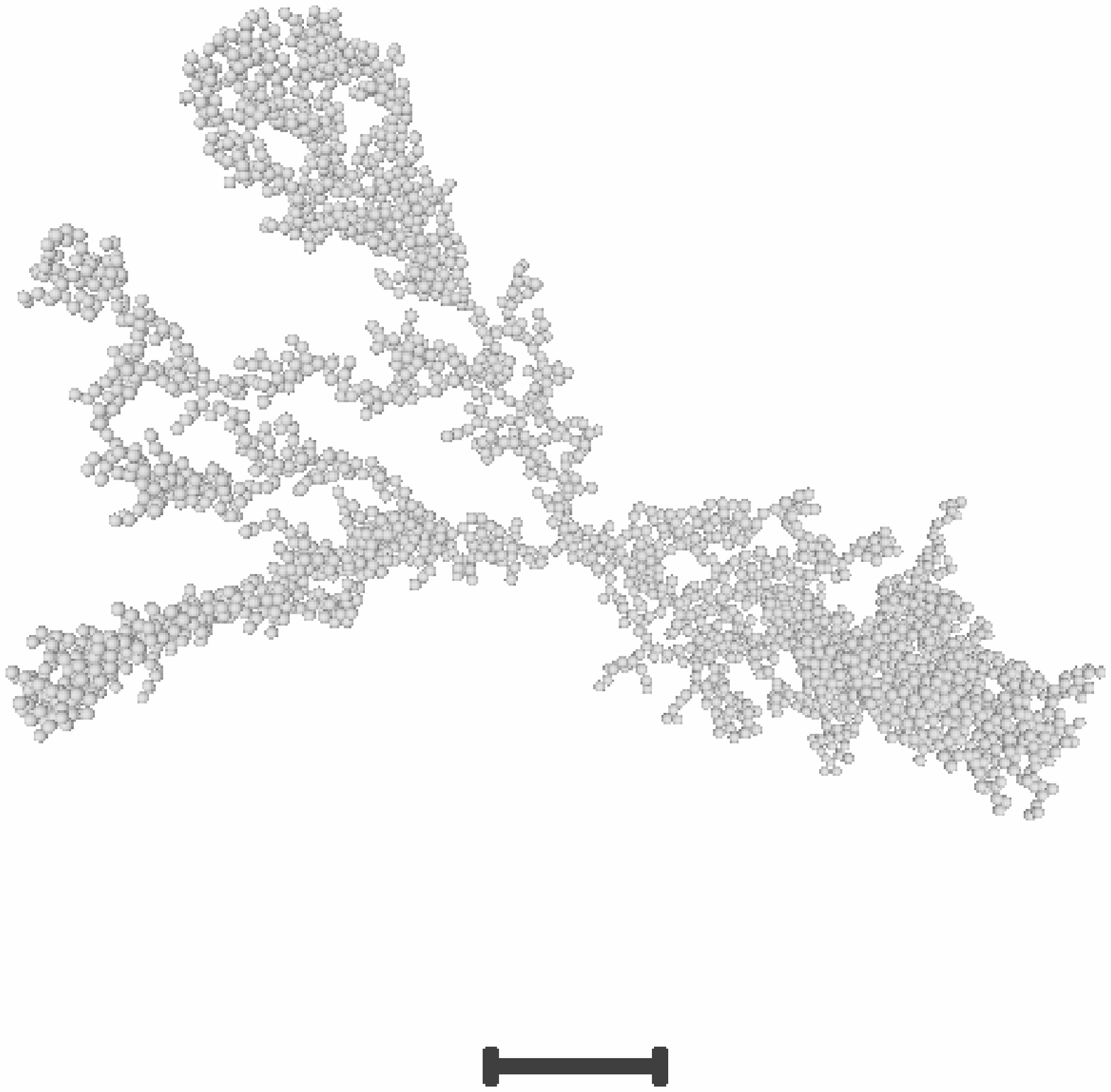}\\
\centerline{$\gamma=6.08$}} &
\parbox{\setw}{\includegraphics[width=\setw]{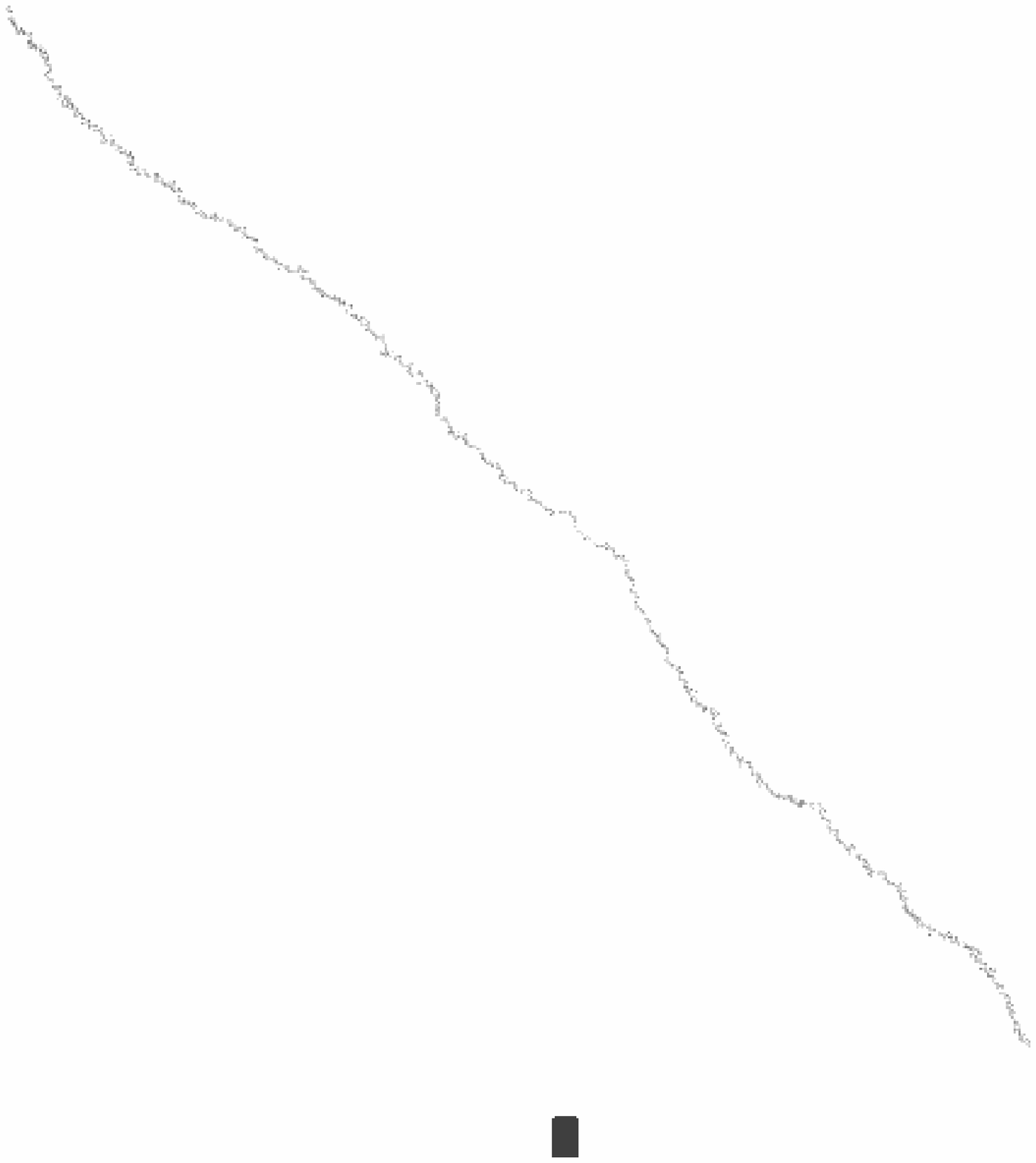}\\
\centerline{$\gamma=53.3$}} \\
\end{tabular}
}
\end{center}
\caption{Pictures of the fractal aggregates we used in our calculations. Indicated by the bar below the aggregates is the diameter of the corresponding volume equivalent sphere.}
\label{fig:Fractals}
\end{figure*}

Note that for a fractal aggregate the circumscribed sphere radius and the volume equivalent radius cannot be the same. The reason is that the constituents are homogeneous spheres, therefore they do not allow for a full packing of the volume of the circumscribed sphere. The maximum packing density of a collection of spheres with the same radius is $\pi/\sqrt{18}$ \citep{Hales1992}. This implies that for an aggregate composed of homogeneous spheres the minimum value of $\gamma=(\sqrt{18}/\pi)^{1/3}\approx 1.105$. The maximum value of $\gamma$ depends on the number of spheres that make up the aggregate. The least dense packing is a straight line of spheres and results in $\gamma=N^{2/3}$.

%For an arbitrary combination of $k_f$, $D_f$ and $N$ the cicumscribed sphere
%radius is given by \citet{Filippov}. From this it is possible to derive
%$\gamma$ for an arbitrary fractal aggregate obeying the scaling law
%(Eq.~\ref{eq:scaling law})
%\begin{equation}
%\label{eq:gamma}
%\gamma=\frac{1}{N^{1/3}}\left(1+
%\sqrt{\frac{N^2}{N-1}\left(\frac{N}{k_f}\right)^{2/D_f}-
%\frac{N}{N-1}-
%N\left(\frac{N-1}{k_f}\right)^{2/D_f}}\right).
%\end{equation}

In nature fractal aggregates can grow with a wide range of fractal dimensions depending on the formation mechanism and the environment in which they form. In astronomical environments the fractal dimension can vary between approximately 1 up to almost 3. Almost linear aggregates ($D_f\approx 1$) might form when the particles have an electric or magnetic dipole and are in an external field \citep{magnetic-I,magnetic-II}. In the case that grain growth occurs by coagulation of aggregates with approximately equal size the resulting aggregates will have a fractal dimension of approximately 1.8 to 2.1 \citep{kempf99}. On the other hand, when the growth occurs by adding single monomers to a larger aggregate, aggregates with a fractal dimension of 3 might form \citep{Ball84}.

In this paper we consider aggregates composed of homogeneous spheres with fractal dimensions ranging from $1.2$ up to $2.8$. The radius of the monomers, $a$, is chosen to be $0.4\,\mu$m. For particles composed of an amorphous silicate, the absolute value of the refractive index in the $10\,\mu$m region reaches values up to $|m|\approx 2.2$ with a corresponding maximum value of $ka|m|=0.5$. Thus, when applying DDA in this spectral region for the material we consider, each monomer can be represented by a single dipole (see Eq.~\ref{eq:DDA}). By using several values of $k_f$ we find that the results are not very sensitive to the fractal prefactor chosen. For all aggregates we use $k_f=2.0$. Pictures of the fractal aggregates we constructed for various numbers of monomers, and thus for various volume equivalent radii, are shown in Fig.~\ref{fig:Fractals}. The bar below the pictures shows the diameter of a volume equivalent sphere. Also denoted are the corresponding values of $\gamma$. Thus we find that for a fractal aggregate with $D_f=1.2$ and a volume equivalent radius $r_V=6\,\mu$m, we have ${r_c=\gamma~r_V\approx 320\,\mu}$m.
%The value of $\gamma$ is computed from the actual structure of the aggregate
%which can be slightly different from that obtained using Eq.~(\ref{eq:gamma})
%for small values of $N$. 

When we consider growth, we do not scale the aggregate but add more grains to it. This is an important difference with the Gaussian random spheres, since it implies that aggregates with different sizes have different shapes. For Gaussian random spheres the size is changed by scaling the entire grain, which implies that the shape of the particle is independent of its size.

\section{Results}
\label{sec:results}

\subsection{The 10\,$\mu$m amorphous silicate feature}

The absorption spectrum of silicate particles display a feature around $10\,\mu$m due to the stretching mode of the Si-O bond in the SiO$_4$ tetrahedra of which silicates are made. Here we will concentrate on the $10\,\mu$m feature of amorphous olivine-type silicate, which is the most important dust constituent in many astronomical environments. The chemical formula of amorphous olivine-type silicate is Mg$_{2x}$Fe$_{2-2x}$SiO$_4$, where $0\le x\ge 1$ gives the iron to magnesium ratio. We use the refractive index as a function of wavelength as measured by \citet{1995A&A...300..503D}, for olivine with $x\approx 0.5$.

The spectral shape and amplitude of the mass absorption coefficient of particles composed of an amorphous silicate as a function of wavelength in the $10\,\mu$m region is sensitive to the size and shape of the grains causing it. Computations using homogeneous spheres show that an increase in particle size tends to broaden and flatten the $10\,\mu$m feature. When Mie theory is employed to compute the emission spectrum, the feature strength diminishes rapidly when the grain size is increased. Using the porous sphere approximation \citet{1990ApJ...361..251H} have shown that this change in feature strength and shape is strongly affected by the porosity factor of the particle. Using approximate methods to compute the emission spectra they showed that the spectra of large, porous spheres resemble those of small compact ones.

\subsection{Fractal aggregates versus Gaussian random spheres}

\begin{figure*}[!t]
\begin{center}
\resizebox{15cm}{!}{\includegraphics{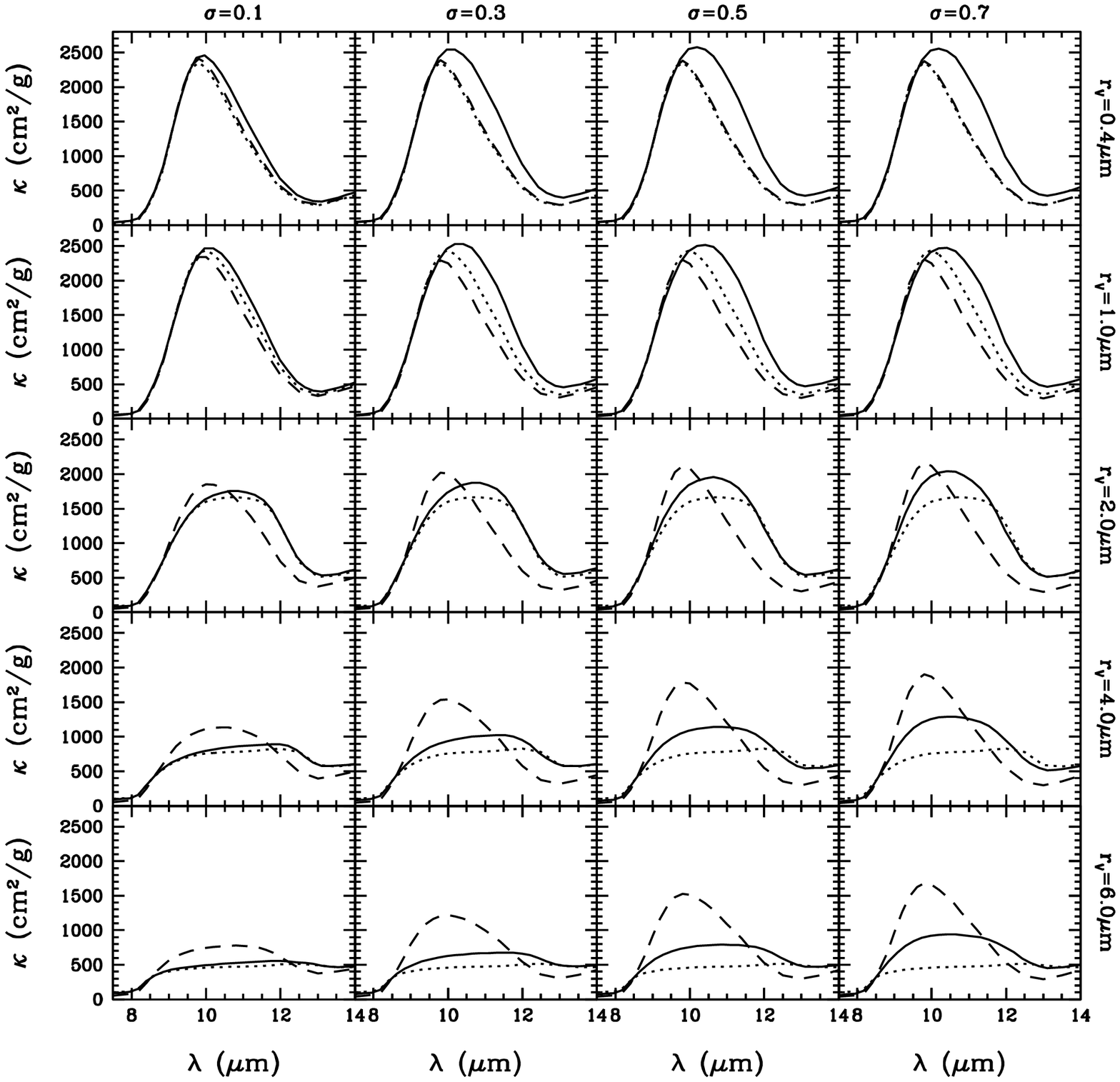}}
\end{center}
\caption{The mass absorption coefficient, $\kappa$, averaged over all particle orientations as a function of wavelength for Gaussian Random Spheres composed of amorphous olivine-type silicate (solid lines) with various values of the shape parameter $\sigma$ and grain sizes $r_V$. Also shown are the spectra for volume equivalent homogeneous spheres (dotted lines) and for equivalent porous spheres with a porosity factor given by Eq.~(\ref{eq:porosity}) (dashed lines). Particle size is increased from top to bottom, while the particle non-sphericity is increased from left to right. Note that in accordance with the shape of the particles, the porosity of the grains increases from left to right. The refractive index of amorphous olivine-type silicate as a function of wavelength used in the calculations was taken from \citet{1995A&A...300..503D}.}
\label{fig:10 micron Gsphere}
\end{figure*}

\begin{figure*}[!t]
\begin{center}
\resizebox{15cm}{!}{\includegraphics{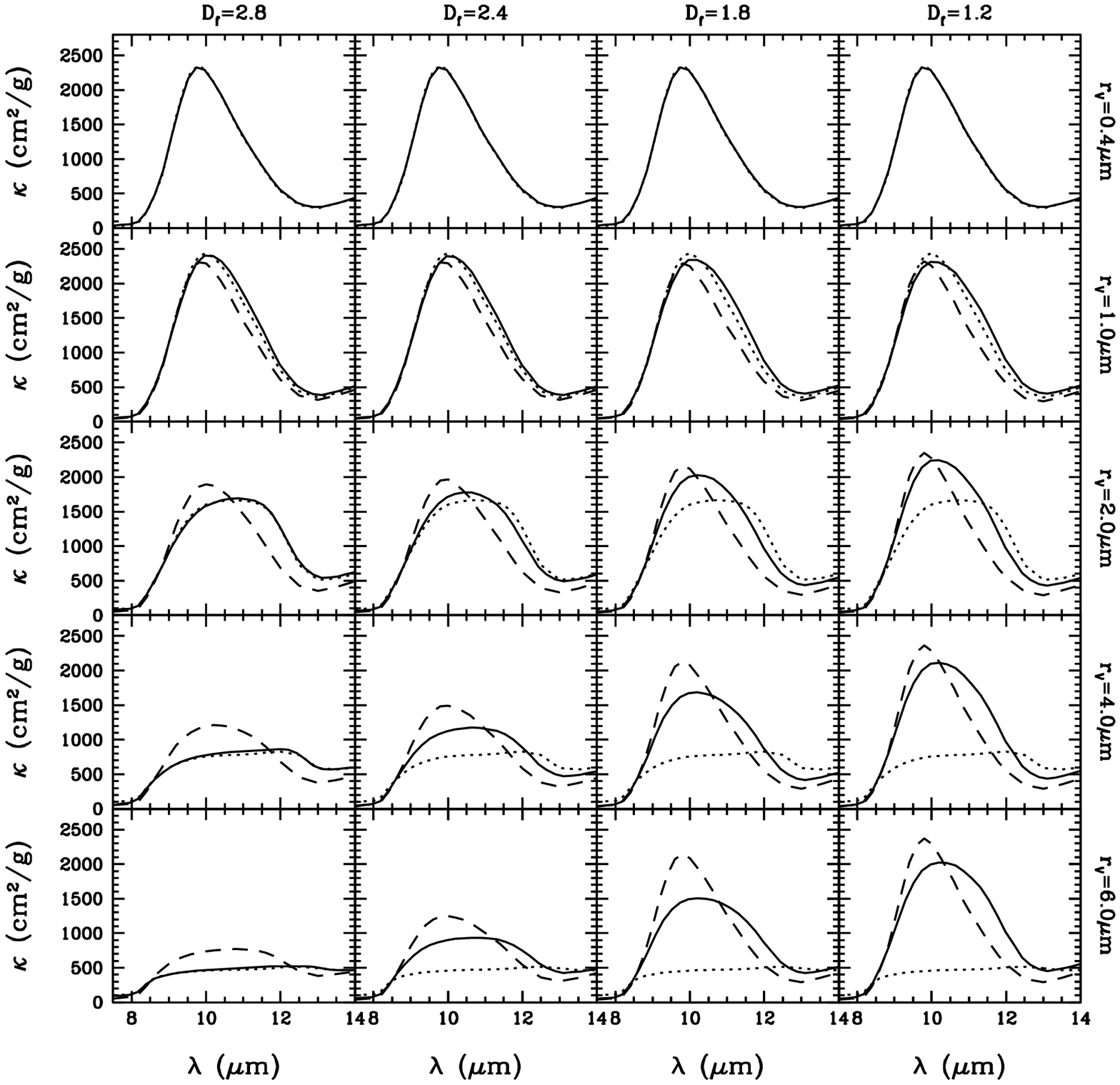}}
\end{center}
\caption{Same as Fig.~\ref{fig:10 micron Gsphere} but for fractal aggregates with various values of the fractal dimension (solid lines). The increase in particle size for aggregates implies an increase of the number of constituents (see Fig.~\ref{fig:Fractals}). Note that in this figure the porosity increases not only from left to right but also from top to bottom (see also Fig.~\ref{fig:Fractals})}
\label{fig:10 micron Fractals}
\end{figure*}

The solid curves in Fig.~\ref{fig:10 micron Gsphere} display the mass absorption coefficients as a function of wavelength for the various Gaussian random spheres (Fig.~\ref{fig:Gaussian}), and different particle sizes. Also shown in this figure is the mass absorption coefficient of volume equivalent homogeneous spherical particles (dotted lines), and the mass absorption coefficient of equivalent porous spheres with porosity factor given by Eq.~(\ref{eq:porosity}) (dashed lines). It is clear that in all cases an increase of the particle size beyond 1\,$\mu$m, leads to a decrease of the strength of the $10\,\mu$m feature. Also, for the grains with a volume equivalent radius larger than 1\,$\mu$m, increasing the value of $\sigma$ increases the peak value. The reason is probably that the bumps and spikes sticking out of the particle for high values of $\sigma$ (see Fig.~\ref{fig:Gaussian}) roughly interact with the radiation like small separate particles. In general, the feature as computed using volume equivalent homogeneous spheres is narrower and weaker. Also, it is clear that the feature computed using the porous sphere approximation is in general narrower and stronger, thus resembling the feature of much smaller particles.
When for homogeneous spheres the grain size is increased, the feature weakens and flattens, i.e. it becomes more flat-topped. For the Gaussian random spheres the feature also weakens when the grain size is increased, however, the shape of the feature still shows a more rounded top.

By considering the spectra of different realizations of the Gaussian random spheres with the same shape parameters $\sigma$ and $\Gamma$ but different seeds for the random number generator we find that the differences in the absorption spectra are small and that all trends are conserved (not shown in the figure).

Plotted by the solid curves in Fig.~\ref{fig:10 micron Fractals} is the mass absorption coefficient as a function of wavelength for the fractal aggregates with different fractal dimensions and aggregate sizes as shown in Fig.~\ref{fig:Fractals}. It is clear from this figure that the $10\,\mu$m feature of fractal aggregates with a low value of the fractal dimension in general has a much weaker dependence on the particle size than that of aggregates with a higher value of the fractal dimension.
For a fractal aggregate with $D_f=1.2$ the feature strength is only very slightly decreased even if the volume equivalent radius is increased from $r_V=0.4\,\mu$m to $6\,\mu$m, in which case $r_c=320\,\mu$m. The reason is probably that the constituents in fractal aggregates with low fractal dimensions are separated by large distances. This reduces the interaction between the different volume elements within the aggregate. Thus, the monomers interact, to a certain extent, as if they were separate small particles, displaying a feature typical for small dust grains. 
We have to note here that there is a difference between the spectra for the
separate constituent (upper panels) and the fully grown aggregates (shown in
the lower panels) even for the most fluffy aggregates. 
This means that even in the most extreme case $D_f=1.2$ there has to be interaction between the constituents of the aggregate. Thus, the effects of aggregation on the spectral signature will always be visible, even in the case that the aggregates form almost linear chains.
In general, when increasing the fractal dimension, the spectral signature of the monomers becomes more apparent in the spectrum of the aggregate.
Comparing the resulting spectra for the fractal aggregates with homogeneous spheres we see that, as with the Gaussian random spheres, the features for the larger fractal aggregates also have a generally more rounded shape. The spectra of very compact aggregates ($D_f=2.8$) show virtually no differences with those of homogeneous spheres.

Also for the fractal aggregates we considered different realizations of the particles with the same shape parameters but different seeds of the random number generator and found that the differences in the spectra obtained are negligible.

Comparing the features obtained using DDA calculations for fractal aggregates with those obtained using the equivalent porous spheres (dashed curves in Fig.~\ref{fig:10 micron Fractals}) it is clear that the effect of fluffiness is overestimated using porous spheres. The features obtained using this approximation are, in general, much sharper, stronger and narrower than those obtained using DDA calculations, thus resembling more the feature of the constituents of the aggregates. The effective medium theory used to compute the spectra of porous spheres assumes that the constituent particles of the aggregate are distributed randomly over the entire volume of the circumscribing sphere. As mentioned before, this is in sharp contrast with realistic aggregates where the constituents have to be in contact, thus increasing the interactions between the aggregate constituents. In this way it is easy to understand that the porous sphere approximation underestimates the effects of grain growth for fluffy aggregates. Our detailed calculations qualitatively confirm the conclusion by \citet{1990ApJ...361..251H} that by increasing particle fluffiness the spectral signature of the constituent particles of the aggregate become more visible. However, we also have to conclude that the porous sphere approximation severely overestimates this effect for realistic aggregates.

\section{Implications for grain growth and dust modeling}
\label{sec:discussion}

In this section we will quantify the effects of the assumption of particle shape on the derived grain size and dust mass when fitting observed infrared 10$\,\mu$m spectra. We do this by taking the 10$\,\mu$m spectra presented in the previous section and fitting these with the 10$\,\mu$m spectrum of an ensemble of homogeneous spheres with radius $r_f$ and total mass $M_f$. In the same way we fitted the spectra using an ensemble of porous spheres, applying the porous sphere approximation, with volume equivalent radius $r_p$ and total mass $M_p$. The porosity of these spheres was taken to be equal to the porosity of the particle whose spectrum was fitted. The fits using homogeneous spheres showed that, in general, the spectral shapes shown in Figs.~\ref{fig:10 micron Fractals} and \ref{fig:10 micron Gsphere} are very well reproduced using homogeneous spheres. In almost all cases the fitted curves deviate less than the linewidth from the curves shown in Figs.~\ref{fig:10 micron Gsphere} and \ref{fig:10 micron Fractals}. Only for the large ($4$ and $6\,\mu$m) Gaussian random spheres with high values of $\sigma$ and the large fractal aggregates with intermediate values of $D_f$ the fits are of a slightly lesser quality. In these cases, the nonspherical particles display a feature with low contrast combined with a relatively rounded top. This cannot be reproduced accurately by using homogeneous spherical particles. The fits using the porous sphere approximation showed also quite good agreement. However, for the large ($r_V>2\,\mu$m), very fluffy aggregates ($D_f=1.2$ and $1.8$), the agreement is much less satisfying.
The fact that most spectra are fitted very well implies that it is difficult to determine particles shapes from the shape of the 10$\,\mu$m feature alone. Thus, one has to resort to other observables, like for example the degree of linear polarization in the visible part of the spectrum, to obtain information on the particle shape \citep[see e.g.][]{2005astro.ph..5603M}. 

\begin{figure}[!t]
\resizebox{\hsize}{!}{\includegraphics{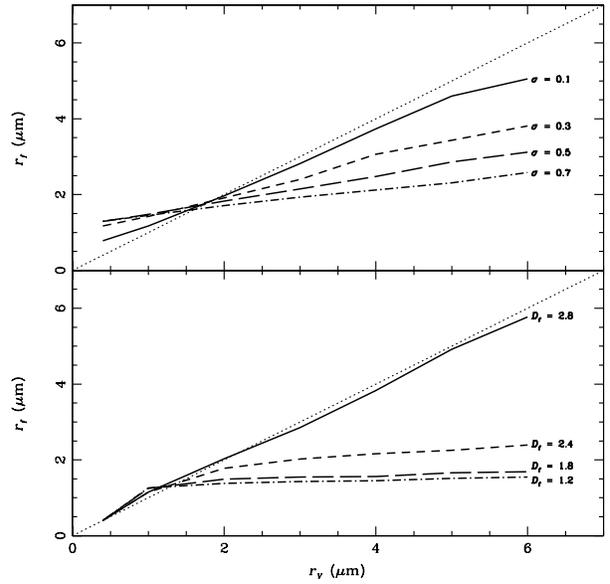}}
\caption{The best fit size of a homogeneous sphere to the 10$\,\mu$m amorphous silicate spectra, $r_f$, of particles with different shapes as a function of the volume equivalent radius of the particles, $r_V$. The top panel is for the Gaussian random spheres with different values of $\sigma$, while the lower panel is for the fractal aggregates with different fractal dimensions, $D_f$. The values of $\sigma$ and $D_f$ are indicated at the right side of the curves. The dotted lines represent the case in which the fitted size and the real size are equal, i.e. $r_V=r_f$.}
\label{fig:fit parameters}
\end{figure}

The resulting values of the best fitting $r_f$ and $r_p$ for the different values of $\sigma$, $D_f$, and $r_V$ are presented in Table~\ref{tab:fit parameters rf} and the corresponding values of $M_f$ and $M_p$ in Table~\ref{tab:fit parameters Mf}. In Fig.~\ref{fig:fit parameters} we plot $r_f$ as a function of $r_V$ for all particle shapes we considered. It is clear that in almost all cases the size of the particles is underestimated when fitting the 10$\,\mu$m absorption spectrum using homogeneous spheres. This is a consequence of the fact that when using homogeneous spheres the particle mass is concentrated in a much smaller volume than when realistically shaped particles are used. This increases the interactions between the volume elements of the particle, leading to a stronger effect of particle size on the 10$\,\mu$m feature when using homogeneous spheres. In contrast, when using equivalent porous spheres to fit the spectra, the particle size is severely overestimated. This in turn is caused by the fact that when using equivalent porous spheres, the mass is distributed over a much larger volume than when using realistic particle shapes, thus leading to a much weaker effect of particle size.
For the fractal aggregates with very low fractal dimension these effects are most extreme. Extrapolation of the curves in Fig.~\ref{fig:fit parameters} beyond $6\,\mu$m for fractal aggregates with low fractal dimension suggests that even for very large aggregates the 10$\,\mu$m feature will display a signature typical for relatively small homogeneous spheres. The mass estimate we get from the fit using homogeneous spheres is in most cases close to the real mass ($M_f$ is relatively close to unity). For the porous spheres, this is slightly worse ($M_p$ is close to $2$ in many cases), but still not very extreme. This is a consequence of the fact that the particles we consider are not extremely large. For very large particles, $r_V\gtrsim 20\,\mu$m, the error in the mass estimate will be much larger. For the intermediate fractal dimensions and for the irregularly shaped compact particles we make an error in the mass estimate of 30\% for the largest values of $r_V$ when using homogeneous spheres to fit the spectra.

In several studies of the $10\,\mu$m spectra of circumstellar disks \citep[see e.g.][]{2001A&A...375..950B, 2004ApJ...610L..49H, 2004Natur.432..479V, 2005A&A...437..189V} particle size effects have been modeled by using a typical size for the large particles of $r_V=1.5$ or $2.0\,\mu$m. The spectral signature of particles with these sizes is reported in these spectra. Since this grain size is significantly larger than that derived for the dust in the interstellar medium \citep{2004ApJ...609..826K} this is interpreted as grain growth.
In these studies the particles are considered to be compact, and in most cases homogeneous spheres. However, models of grain growth in astronomical environments show that aggregates with a wide range of fractal dimensions can be formed. As we showed, in general, fractal aggregates and irregularly shaped compact particles display the spectral signature of smaller homogeneous spheres. Thus, when interpreting emission spectra of astronomical objects using compact particles, the size of the emitting grains is generally underestimated.
A good estimate of the real particle shape or fractal dimension may be available from, for example, theoretical arguments. In that case the $r_f$ derived from fitting 10$\,\mu$m spectra of astronomical sources using homogeneous spheres can be used, in combination with Table~\ref{tab:fit parameters rf}, to estimate the volume equivalent radius of the particles, $r_V$.

\begin{table*}[!t]
\begin{center}
\begin{tabular}{c|cccc|cccc|cccc|cccc|}
		& \multicolumn{8}{c|}{Homogeneous Spheres}&\multicolumn{8}{c|}{Porous sphere approximation}\\
		& \multicolumn{4}{c|}{\footnotesize Gaussian random spheres}&\multicolumn{4}{c|}{\footnotesize Fractal aggregates}
		& \multicolumn{4}{c|}{\footnotesize Gaussian random spheres}&\multicolumn{4}{c|}{\footnotesize Fractal aggregates} \\
		&	$\sigma$&	$\sigma$&	$\sigma$&	$\sigma$&	$D_f$	&	$D_f$	&	$D_f$	&	$D_f$
		&	$\sigma$&	$\sigma$&	$\sigma$&	$\sigma$&	$D_f$	&	$D_f$	&	$D_f$	&	$D_f$\\
		&	$0.1$&	$0.3$&	$0.5$&	$0.7$&	$2.8$	&	$2.4$	&	$1.8$	&	$1.2$
		&	$0.1$&	$0.3$&	$0.5$&	$0.7$&	$2.8$	&	$2.4$	&	$1.8$	&	$1.2$\\
\hline
$r_V=0.4\,\mu$m	&	$0.8$   &	  $1.2$   &     $1.3$	&	$1.3$   &	  $0.4$   &     $0.4$	&	$0.4$   &	  $0.4$ &	  $1.0$   &     $2.9$	&	$5.8$   &	  $8.0$   &     $0.4$	&	$0.4$   &	  $0.4$   &     $0.4$\\
$r_V=1.0\,\mu$m	&	$1.2$   &	  $1.4$   &     $1.5$	&	$1.5$   &	  $1.2$   &     $1.1$	&	$1.3$   &	  $1.3$ &	  $1.2$   &     $4.5$	&	$8.6$   &	  $11.1$  &     $2.1$	&	$2.1$   &	  $3.8$   &     $12.8$\\
$r_V=2.0\,\mu$m	&	$2.0$   &	  $1.9$   &     $1.8$	&	$1.7$   &	  $2.0$   &     $1.8$	&	$1.5$   &	  $1.4$ &	  $3.9$   &     $9.9$	&	$15.8$  &	  $17.9$  &     $5.1$	&	$5.8$   &	  $13.0$  &     $122.2$\\
$r_V=4.0\,\mu$m	&	$3.7$   &	  $3.1$   &     $2.5$	&	$2.1$   &	  $3.8$   &     $2.2$	&	$1.6$   &	  $1.4$ &	  $9.2$   &     $21.6$  &	$30.1$  &	  $31.2$  &     $11.8$  &	$11.6$  &	  $36.9$  &     $1101.7$\\
$r_V=6.0\,\mu$m	&	$5.1$   &	  $3.8$   &     $3.1$	&	$2.6$   &	  $5.8$   &     $2.4$	&	$1.7$   &	  $1.5$ &	  $13.6$  &     $30.4$  &	$41.6$  &	  $40.1$  &     $14.6$  &	$16.4$  &	  $76.2$  &     $4391.8$\\
\hline
\end{tabular}
\end{center}
\caption{The radius of the best fit homogeneous sphere, $r_f$ (left columns), and the radius of the best fit porous sphere, $r_p$ (right columns), in $\mu$m for the absorption spectra of the Gaussian random spheres and the fractal aggregates.}
\label{tab:fit parameters rf}
\end{table*}

\begin{table*}[!t]
\begin{center}
\begin{tabular}{c|cccc|cccc|cccc|cccc|}
		& \multicolumn{8}{c|}{Homogeneous Spheres}&\multicolumn{8}{c|}{Porous sphere approximation}\\
		& \multicolumn{4}{c|}{\footnotesize Gaussian random spheres}&\multicolumn{4}{c|}{\footnotesize Fractal aggregates}
		& \multicolumn{4}{c|}{\footnotesize Gaussian random spheres}&\multicolumn{4}{c|}{\footnotesize Fractal aggregates} \\
		&	$\sigma$&	$\sigma$&	$\sigma$&	$\sigma$&	$D_f$	&	$D_f$	&	$D_f$	&	$D_f$
		&	$\sigma$&	$\sigma$&	$\sigma$&	$\sigma$&	$D_f$	&	$D_f$	&	$D_f$	&	$D_f$\\
		&	$0.1$&	$0.3$&	$0.5$&	$0.7$&	$2.8$	&	$2.4$	&	$1.8$	&	$1.2$
		&	$0.1$&	$0.3$&	$0.5$&	$0.7$&	$2.8$	&	$2.4$	&	$1.8$	&	$1.2$\\
\hline
$r_V=0.4\,\mu$m	&	$1.0$   &	  $1.1$   &     $1.1$	&	$1.1$   &	  $1.0$   &     $1.0$	&	$1.0$   &	  $1.0$ &	  $1.0$   &     $1.4$	&	$1.7$   &	  $1.7$   &     $1.0$	&  $1.0$	  &	  $1.0$   &     $1.0$\\
$r_V=1.0\,\mu$m	&	$1.0$   &	  $1.1$   &     $1.2$	&	$1.2$   &	  $1.0$   &     $1.0$	&	$1.0$   &	  $1.0$ &	  $1.2$   &     $1.7$	&	$1.9$   &	  $1.9$   &     $1.2$	&  $1.2$	  &	  $1.4$   &     $1.6$\\
$r_V=2.0\,\mu$m	&	$1.0$   &	  $1.0$   &     $1.1$	&	$1.1$   &	  $1.0$   &     $1.0$	&	$1.0$   &	  $1.0$ &	  $1.5$   &     $2.1$	&	$2.2$   &	  $2.1$   &     $1.7$	&  $1.6$	  &	  $1.7$   &     $1.8$\\
$r_V=4.0\,\mu$m	&	$1.0$   &	  $0.9$   &     $0.8$	&	$0.8$   &	  $1.0$   &     $0.8$	&	$0.8$   &	  $1.0$ &	  $1.7$   &     $2.4$	&	$2.3$   &	  $2.1$   &     $1.9$	&  $1.7$	  &	  $1.7$   &     $1.8$\\
$r_V=6.0\,\mu$m	&	$0.9$   &	  $0.8$   &     $0.7$	&	$0.7$   &	  $1.0$   &     $0.7$	&	$0.8$   &	  $1.0$ &	  $1.7$   &     $2.2$	&	$2.2$   &	  $1.9$   &     $1.7$	&  $1.6$	  &	  $1.8$   &     $1.9$\\
\hline
\end{tabular}
\end{center}
\caption{The mass scaling factor of the best fit homogeneous sphere, $M_f$ (left columns), and of the best fit porous sphere, $M_p$ (right columns), for the absorption spectra of the Gaussian random spheres and the fractal aggregates.}
\label{tab:fit parameters Mf}
\end{table*}

For example, say that we derive from the shape of an observed 10$\,\mu$m feature using homogeneous spheres that the typical particle size $r_f=1.7\,\mu$m. Furthermore, we expect that the grains present in the environment we are observing are fractal aggregates with $D_f\approx 1.8$. We can then estimate from Table~\ref{tab:fit parameters rf} that the true volume equivalent radius of the grains is $r_V\approx 6\,\mu$m. The linear extent of the aggregates can be estimated using $\gamma$ to be $r_c\approx 36\,\mu$m (see Fig.~\ref{fig:Fractals}). In addition, from Table~\ref{tab:fit parameters Mf} it can be found that the derived dust mass is a likely underestimate of the real dust mass by roughly 20\%.

Since the curves for the low fractal dimension are very flat for $r_V\gtrsim 1\,\mu$m, the estimate of $r_V$ strongly depends on the accuracy of $r_f$. Furthermore, the fractal dimension that is employed can change the estimated value of $r_V$ from a given value of $r_f$ significantly. Coagulation calculations and experiments play a crucial role in determining the possible shapes and fractal dimensions that can be formed under different conditions. Therefore, these calculations and experiments are very important for a correct analysis of 10$\,\mu$m spectra.

Models of grain growth predict that, for example, in protoplanetary disks, grain growth rapidly removes small grains \citep{Dullemond2005}. When homogeneous spheres are used to compute the $10\,\mu$m spectra of amorphous silicate particles, already for a grain with a $4\,\mu$m radius the feature is flattened so much that it becomes very hard to detect. This contradicts observations of protoplanetary disks where prominent $10\,\mu$m features are found \citep[see e.g.][]{2001A&A...375..950B}. \citet{Dullemond2005} propose a possible explanation for this discrepancy by considering destructive particle collisions. This causes an equilibrium situation with a significant amount of small grains. In addition to this, we have shown that when the particles grow as fluffy aggregates, the $10\,\mu$m feature will be visible for much larger particle sizes.

A flattened, square $10\,\mu$m feature as seen in the absorption spectra of homogeneous spheres is observed in the $10\,\mu$m emission spectra of some protoplanetary disks \citep[see e.g][]{2001A&A...375..950B}. Since the features of very irregularly shaped particles or fluffy aggregates as presented in Figs.~\ref{fig:10 micron Gsphere} and \ref{fig:10 micron Fractals} show a somewhat more rounded top, one could argue that this suggests that the particles in these protoplanetary disks may be relatively compact. However, a firmer analysis of the observations is needed in order to confirm this.

\section{Conclusions}
\label{sec:conclusions}

We have presented calculations of the $10\,\mu$m absorption spectra of complex shaped compact particles and fractal aggregates with various fractal dimensions. We have compared the resulting spectra and studied the dependence of the spectral signature on the size of the particles. It is clear that the size dependence of the spectral signature of fractal aggregates depends strongly on the fractal dimension of the aggregate. For very fluffy aggregates, i.e. low fractal dimensions, the spectral signature of very large aggregates still looks like that of very small compact particles. We have investigated if the absorption spectra of complex particles can be approximated by those obtained using volume equivalent porous spheres with the same volume filling factor as proposed by \citet{1990ApJ...361..251H}. We conclude that the interactions between the volume elements making up the particle are underestimated when we use this porous sphere approximation, applying Garnett effective medium theory, thus underestimating the effects of particle size.

For the complex shaped compact particles the size dependence of the spectral signature also depends on the actual shape of the emitting grains. In general we find that the strength of the $10\,\mu$m absorption spectrum of a nonspherical particle is equal to that of a smaller homogeneous spherical particle. 

Observed $10\,\mu$m emission and absorption spectra are often interpreted using compact, and in most cases homogeneous spherical, particles. We show that this leads to an underestimate of the actual grain size when the emitting grains are either nonspherical compact particles or fractal aggregates. We present a way of estimating the true particle size and mass, when an estimate of the shape or fractal dimension of the particles can be provided. We also show that when equivalent porous spheres are used to fit observed 10$\,\mu$m spectra, the particle sizes are severely overestimated.

Analysis of the $10\,\mu$m emission spectra of circumstellar disks show that the spectral signature of homogeneous spherical particles with radii of approximately $2\,\mu$m is present in these disks. When the emitting grains are actually, for instance, fractal aggregates with a fractal dimension of approximately $1.8$ this implies a volume equivalent radius of these aggregates of at least $r_V=6\,\mu$m. The radius of the circumscribed sphere is then even of the order of $r_c\approx 36\,\mu$m. In environments where aggregates with even lower fractal dimensions can grow, grain sizes will be even more severely underestimated when using compact particles.

\begin{acknowledgements}
We are grateful to the referee, N.~V. Voshchinnikov, for valuable comments on an earlier version of this paper.
\end{acknowledgements}


\begin{thebibliography}{34}
\expandafter\ifx\csname natexlab\endcsname\relax\def\natexlab#1{#1}\fi

\bibitem[{Ball \& Witten(1984)}]{Ball84}
Ball, R.~C. \& Witten, T.~A. 1984, \pra, 29, 2966

\bibitem[{{Bouwman} {et~al.}(2001){Bouwman}, {Meeus}, {de Koter}, {Hony},
  {Dominik}, \& {Waters}}]{2001A&A...375..950B}
{Bouwman}, J., {Meeus}, G., {de Koter}, A., {et~al.} 2001, \aap, 375, 950

\bibitem[{{Dominik} \& {N{\" u}bold}(2002)}]{magnetic-I}
{Dominik}, C. \& {N{\" u}bold}, H. 2002, Icarus, 157, 173

\bibitem[{{Dorschner} {et~al.}(1995){Dorschner}, {Begemann}, {Henning}, {J{\"
  a}ger}, \& {Mutschke}}]{1995A&A...300..503D}
{Dorschner}, J., {Begemann}, B., {Henning}, T., {J{\" a}ger}, C., \&
  {Mutschke}, H. 1995, \aap, 300, 503

\bibitem[{Draine(2000)}]{Draine2000}
Draine, B.~T. 2000, in Light Scattering by Nonspherical Particles: Theory,
  Measurements, and Applications, ed. M.~I. Mishchenko, J.~W. Hovenier, \&
  L.~D. Travis (San Diego: Academic Press), 131--145

\bibitem[{{Draine} \& {Flatau}(1994)}]{1994OSAJ...11.1491D}
{Draine}, B.~T. \& {Flatau}, P.~J. 1994, Optical Society of America Journal A,
  11, 1491

\bibitem[{{Draine} \& {Goodman}(1993)}]{1993ApJ...405..685D}
{Draine}, B.~T. \& {Goodman}, J. 1993, \apj, 405, 685

\bibitem[{{Dullemond} \& {Dominik}(2005)}]{Dullemond2005}
{Dullemond}, C.~P. \& {Dominik}, C. 2005, \aap~in press

\bibitem[{{Fabian} {et~al.}(2001){Fabian}, {Henning}, {J{\" a}ger}, {Mutschke},
  {Dorschner}, \& {Wehrhan}}]{2001A&A...378..228F}
{Fabian}, D., {Henning}, T., {J{\" a}ger}, C., {et~al.} 2001, \aap, 378, 228

\bibitem[{{Filippov} {et~al.}(2000){Filippov}, {Zurita}, \&
  {Rosner}}]{Filippov}
{Filippov}, A.~V., {Zurita}, M., \& {Rosner}, D.~E. 2000, Journal of Colloid
  and Interface Science, 229, 261

\bibitem[{{Goodman} {et~al.}(1991){Goodman}, {Draine}, \&
  {Flatau}}]{Goodman1991}
{Goodman}, J.~J., {Draine}, B.~T., \& {Flatau}, P.~J. 1991, Optics Letters, 16,
  1198

\bibitem[{{Hage} \& {Greenberg}(1990)}]{1990ApJ...361..251H}
{Hage}, J.~I. \& {Greenberg}, J.~M. 1990, \apj, 361, 251

\bibitem[{{Hales}(1992)}]{Hales1992}
{Hales}, T.~C. 1992, Journal of Computational and Applied Mathematics, 44, 41

\bibitem[{{Hansen} \& {Hovenier}(1974)}]{1974JAtS...31.1137H}
{Hansen}, J.~E. \& {Hovenier}, J.~W. 1974, Journal of Atmospheric Sciences, 31,
  1137

\bibitem[{{Henning} \& {Stognienko}(1993)}]{1993A&A...280..609H}
{Henning}, T. \& {Stognienko}, R. 1993, \aap, 280, 609

\bibitem[{Hoekstra {et~al.}(1998)Hoekstra, Grimminck, \& Sloot}]{Hoekstra}
Hoekstra, A.~G., Grimminck, M.~D., \& Sloot, P. M.~A. 1998, International
  Journal of Modern Physics, 9, 87

\bibitem[{{Honda} {et~al.}(2004){Honda}, {Kataza}, {Okamoto}, {Miyata},
  {Yamashita}, {Sako}, {Fujiyoshi}, {Ito}, {Okada}, {Sakon}, \&
  {Onaka}}]{2004ApJ...610L..49H}
{Honda}, M., {Kataza}, H., {Okamoto}, Y.~K., {et~al.} 2004, \apjl, 610, L49

\bibitem[{{Kemper} {et~al.}(2004){Kemper}, {Vriend}, \&
  {Tielens}}]{2004ApJ...609..826K}
{Kemper}, F., {Vriend}, W.~J., \& {Tielens}, A.~G.~G.~M. 2004, \apj, 609, 826

\bibitem[{{Kempf} {et~al.}(1999){Kempf}, {Pfalzner}, \& {Henning}}]{kempf99}
{Kempf}, S., {Pfalzner}, S., \& {Henning}, T.~K. 1999, Icarus, 141, 388

\bibitem[{{Lakhtakia}(1992)}]{1992ApJ...394..494L}
{Lakhtakia}, A. 1992, \apj, 394, 494

\bibitem[{{Lakhtakia} \& {Mulholland}(1993)}]{Lakhtakia1993}
{Lakhtakia}, A. \& {Mulholland}, G.~W. 1993, Journal of Research of the
  National Institute of Standards and Technology, 98, 699

\bibitem[{{Maxwell-Garnett}(1904)}]{1904RSPTA.203..385M}
{Maxwell-Garnett}, J.~C. 1904, Royal Society of London Philosophical
  Transactions Series A, 203, 385

\bibitem[{Mie(1908)}]{Mie}
Mie, G. 1908, Ann Phys., 25, 377

\bibitem[{{Min} {et~al.}(2003){Min}, {Hovenier}, \& {de
  Koter}}]{2003A&A...404...35M}
{Min}, M., {Hovenier}, J.~W., \& {de Koter}, A. 2003, \aap, 404, 35

\bibitem[{{Min} {et~al.}(2005{\natexlab{a}}){Min}, {Hovenier}, \& {de
  Koter}}]{MinHollow}
---. 2005{\natexlab{a}}, \aap, 432, 909

\bibitem[{{Min} {et~al.}(2005{\natexlab{b}}){Min}, {Hovenier}, {de Koter},
  {Waters}, \& {Dominik}}]{2005astro.ph..5603M}
{Min}, M., {Hovenier}, J.~W., {de Koter}, A., {Waters}, L.~B.~F.~M., \&
  {Dominik}, C. 2005{\natexlab{b}}, Icarus, in press

\bibitem[{Mishchenko {et~al.}(2000)Mishchenko, Hovenier, \&
  Travis}]{MishHoveTravis}
Mishchenko, M.~I., Hovenier, J.~W., \& Travis, L.~D., eds. 2000, Light
  Scattering by Nonspherical Particles, Theory, Measurements and Applications
  (San Diego: Academic Press)

\bibitem[{{Muinonen} {et~al.}(1996){Muinonen}, {Nousiainen}, {Fast}, {Lumme},
  \& {Peltoniemi}}]{1996JQSRT..55..577M}
{Muinonen}, K., {Nousiainen}, T., {Fast}, P., {Lumme}, K., \& {Peltoniemi}, J.
  1996, Journal of Quantitative Spectroscopy and Radiative Transfer, 55, 577

\bibitem[{{N{\" u}bold} {et~al.}(2003){N{\" u}bold}, {Poppe}, {Rost},
  {Dominik}, \& {Glassmeier}}]{magnetic-II}
{N{\" u}bold}, H., {Poppe}, T., {Rost}, M., {Dominik}, C., \& {Glassmeier}, K.
  2003, Icarus, 165, 195

\bibitem[{{Purcell} \& {Pennypacker}(1973)}]{1973ApJ...186..705P}
{Purcell}, E.~M. \& {Pennypacker}, C.~R. 1973, \apj, 186, 705

\bibitem[{{Stognienko} {et~al.}(1995){Stognienko}, {Henning}, \&
  {Ossenkopf}}]{1995A&A...296..797S}
{Stognienko}, R., {Henning}, T., \& {Ossenkopf}, V. 1995, \aap, 296, 797

\bibitem[{{van Boekel} {et~al.}(2004){van Boekel}, {Min}, {Leinert}, {Waters},
  {Richichi}, {Chesneau}, {Dominik}, {Jaffe}, {Dutrey}, {Graser}, {Henning},
  {de Jong}, {K{\" o}hler}, {de Koter}, {Lopez}, {Malbet}, {Morel}, {Paresce},
  {Perrin}, {Preibisch}, {Przygodda}, {Sch{\" o}ller}, \&
  {Wittkowski}}]{2004Natur.432..479V}
{van Boekel}, R., {Min}, M., {Leinert}, C., {et~al.} 2004, \nat, 432, 479

\bibitem[{{van Boekel} {et~al.}(2005){van Boekel}, {Min}, {Waters}, {de Koter},
  {Dominik}, {van den Ancker}, \& {Bouwman}}]{2005A&A...437..189V}
{van Boekel}, R., {Min}, M., {Waters}, L.~B.~F.~M., {et~al.} 2005, \aap, 437,
  189

\bibitem[{{Voshchinnikov} {et~al.}(2005){Voshchinnikov}, {Il'in}, \&
  {Henning}}]{2005A&A...429..371V}
{Voshchinnikov}, N.~V., {Il'in}, V.~B., \& {Henning}, T. 2005, \aap, 429, 371

\end{thebibliography}
\end{document}